\documentclass[final]{elsarticle}

\usepackage{amssymb}

\usepackage{booktabs}
\usepackage{graphicx}
\usepackage{adjustbox}
\usepackage{multirow}
\usepackage[table,xcdraw]{xcolor}
\usepackage{url}

\usepackage{graphicx} 
\usepackage{subfigure}
\usepackage{amsmath}
\usepackage[noend]{algpseudocode}
\usepackage{algorithmicx,algorithm}
\usepackage{float}
\usepackage{booktabs}
\usepackage{multirow}
\usepackage{hyperref}
\usepackage{lineno}
\usepackage{graphicx}
\usepackage{amsmath}
\usepackage{amssymb}
\usepackage{booktabs}
\usepackage{booktabs}
\usepackage{multirow}
\usepackage[table,xcdraw]{xcolor}
\usepackage{booktabs}
\usepackage{multirow}
\usepackage{graphicx}
\usepackage[normalem]{ulem}
\useunder{\uline}{\ul}{}
\usepackage{cleveref}
\usepackage{xcolor}

\usepackage{xcolor}

\newcommand{\hyspr}[1]{\textcolor{black}{#1}}

\newcommand{\eg}{\textit{e.g.}}

\newcommand{\ie}{\textit{i.e.}}
\newcommand{\etal}{\textit{et al}}
\newcommand{\hhys}[1]{\textcolor{black}{#1}}
\newcommand{\hhhys}[1]{\textcolor{black}{#1}}
\newcommand{\pr}[1]{\textcolor{black}{#1}}
\newcommand{\prr}[1]{\textcolor{black}{#1}}
\newcommand{\prrr}[1]{\textcolor{black}{#1}}
\newcommand{\revision}[1]{\textcolor{black}{#1}}

\journal{Pattern Recognition}

\begin{document}
\begin{frontmatter}



\title{Texture and Noise Dual Adaptation for Infrared Image Super-Resolution}


\author[label1]{Yongsong Huang}
\author[label1]{Tomo Miyazaki}
\author[label2]{Xiaofeng Liu}
\author[label2]{Yafei Dong}
\author[label1]{Shinichiro Omachi}

\affiliation[label1]{organization={Graduate School of Engineering, Tohoku University},
            city={Sendai},
            postcode={9808579}, 
            country={Japan}}

\affiliation[label2]{organization={Department of Radiology and Biomedical Imaging, Yale University},
            city={New Haven},
            postcode={06519}, 
            country={USA}}

\begin{abstract}

Recent efforts have explored leveraging visible light images to enrich texture details in infrared (IR) super-resolution. However, this direct adaptation approach often becomes a double-edged sword, as it improves texture at the cost of introducing noise and blurring artifacts. \prr{Such imperfections are inherent in the spatial domain of visible images and are accentuated during the imaging process. Enhancing IR image quality by integrating rich texture details from visible images, while minimizing noise transfer, presents a challenging research avenue.} To address these challenges, we propose the \revision{Texture and Noise Dual Adaptation SRGAN (DASRGAN)}, an innovative framework specifically engineered for robust IR super-resolution model adaptation. DASRGAN operates on the synergy of two key components: 1) Texture-Oriented Adaptation (TOA) to refine texture details meticulously, and 2) Noise-Oriented Adaptation (NOA), dedicated to minimizing noise transfer. Specifically, TOA uniquely integrates a specialized discriminator, incorporating a prior extraction branch, and employs a Sobel-guided adversarial loss to align texture distributions effectively. Concurrently, NOA utilizes a noise adversarial loss to distinctly separate the generative and Gaussian noise pattern distributions during adversarial training. Our extensive experiments confirm DASRGAN's superiority. Comparative analyses against leading methods across multiple benchmarks and upsampling factors reveal that DASRGAN sets new state-of-the-art performance standards. Code are available at \url{https://github.com/yongsongH/DASRGAN}.

\end{abstract}



\begin{highlights}

\item  Our contributions focus on the development of DASRGAN, a dedicated framework for super-resolution in infrared images by leveraging visible light images to enrich texture details. This framework is distinguished by its incorporation of Texture-Oriented Adaptation and Noise-Oriented Adaptation. 
\item For Texture-Oriented Adaptation (TOA): We introduce a specialized discriminator, $\mathbb{D}_{s_{\text {trans }}}$, equipped with a Sobel-prior extraction branch, sharpens focus on texture details during adversarial training. Additionally, we propose a novel Texture-Oriented Prior Adversarial Loss, $\mathcal{L}_{\text {trans }}$, formulated to guide the generator network $\mathbb{G}$ in the meticulous alignment of texture patterns. 
\item For Noise-Oriented Adaptation (NOA): We unveil a new Noise-Oriented Adversarial Loss, $\mathcal{L}_n^{\mathbb{G}}$, explicitly engineered to direct the generator network in attenuating the transference of noise between the visible and infrared domains.
\item Our experiments show that DASRGAN significantly surpasses existing methods, setting a new standard in infrared image super-resolution tasks both qualitatively and quantitatively.

\end{highlights}

\begin{keyword}

Super-resolution; Infrared imaging;  Domain adaptation; Deep learning

\end{keyword}

\end{frontmatter}


\section{Introduction\label{sec.1}}


IR imaging plays a pivotal role in a wide range of applications, including medical diagnostics~\cite{Lukose2021OpticalTF}, target detection~\cite{liu2023combining}, remote sensing~\cite{wei2023adversarial}, and autonomous driving~\cite{gupta2021toward}. It offers unique pattern information otherwise unattainable through other means. However, generating high-quality IR images remains a formidable challenge, largely due to constraints in imaging systems and adverse conditions, such as limited line-of-sight and atmospheric fog~\cite{huang2022infrared,huang2021infrared}. In response to this, there has been growing interest in single-image super-resolution (SISR) reconstruction~\cite{wang2020deep,zhao2024ssir}, which aims to recover a high-resolution (HR) IR image from its low-resolution (LR) counterpart. 

In contrast to visible light images, SR tasks for IR images face the challenge of limited sample availability. This scarcity is primarily attributed to the high cost of IR cameras, which necessitate larger-aperture optical components and more sensitive sensors~\cite{huang2022infrared,he2018cascaded}. As a result, researchers often struggle to acquire a sufficient number of training samples for model development. In response, a new paradigm has emerged in IR image super-resolution, namely domain adaptation or transfer learning, where visible light images are utilized to enhance the details of IR images during the fine-tuning phase~\cite{huang2021infrared,honda2019multi}. Visible light images, being more readily available and rich in textural details, provide valuable pattern information that can significantly improve the quality of IR images~\cite{wang2023m}. Two distinct training adaptation paradigms are observed: the Hybrid and the Decoupled models. 1) The Hybrid approach employs visible samples during the pre-training phase to offset the scarcity of IR samples, albeit at the cost of entangled feature spaces for both image types~\cite{honda2019multi,gupta2021toward}. 2) The Decoupled model by Huang \etal~\cite{huang2021infrared} leverages a dual-branch architecture for independent feature extraction from visible and IR samples, achieving pattern decoupling in the feature space. Nonetheless, the approach fails to differentiate between positive texture and negative noise features. Previous approaches~\cite{song2019multimodal} to feature domain adaptation can inadvertently transfer these degradation features from the visible light domain to the IR domain, further compromising the quality of the SR images.

\hyspr{Recent advances in infrared image super-resolution and domain adaptation have significantly improved the ability to handle diverse image artifacts, including noise and texture inconsistencies. For instance, Jiang \etal~\cite{jiang2022deep}
introduced a deep image denoising framework that leverages adaptive priors to enhance denoising performance by learning noise characteristics and effectively removing them from images. Similarly, in few-shot learning approaches, methods such as Few-Shot Learning for Image Denoising (FSLID) have demonstrated robust denoising capabilities with minimal training data by inducing prior features through multi-scale feature recursion~\cite{jiang2023few}. These techniques highlight the importance of leveraging contextual information and adaptive feature extraction to address noise in image domains with varying levels of artifacts.}

Considering the analyses above, we introduce a novel training framework,  Texture and Noise Dual Adaptation SRGAN (DASRGAN), specifically crafted for SR tasks in IR imaging. \prrr{This framework is characterized by its bifocal adaptation mechanism aimed at both enhancing texture fidelity and mitigating noise. In this work, our primary contributions are threefold: \textbf{\textit{1) Texture-Oriented Adaptation (TOA):}} We introduce a specialized discriminator, $\mathbb{D}_{s_{\text {trans }}}$, equipped with a Sobel-prior extraction branch, sharpens focus on texture details during adversarial training. Additionally, we propose a novel Texture-Oriented Prior Adversarial Loss, $\mathcal{L}_{\text {trans }}$, formulated to guide the generator network $\mathbb{G}$ in the meticulous alignment of texture patterns. \textbf{\textit{2) Noise-Oriented Adaptation (NOA):}} We unveil a new Noise-Oriented Adversarial Loss, $\mathcal{L}_n^{\mathbb{G}}$, explicitly engineered to direct the generator network in attenuating the transference of noise between the visible and IR domains. \textbf{\textit{3)}} Our experiments show that DASRGAN significantly surpasses existing methods, setting a new standard in IR image SR tasks both qualitatively and quantitatively.}



\section{Related Work\label{sec.related work}}

\subsection{Deep Networks for IR Image SR}
With the advent of deep learning, neural networks have shown extraordinary capabilities in SR tasks, leading to increased interest in their application for IR image enhancement~\cite{suarez2024enhancement}. A key focus in IR image super-resolution is the reconstruction of edge details, which are inherently more challenging to capture in IR images compared to visible light images. This challenge stems from the unique physical models~\cite{yu1998infrared} that govern the detection of thermal radiation in IR imaging and the synthesis of IR images~\cite{nandhakumar1994unified}. In IR images, high-frequency and low-frequency information typically represent image outlines and edges, respectively. However, edge information in IR images is generally sparser than in visible light images, rendering the task more challenging~\cite{10314035,jiang2023few}. 

The challenge of utilizing deep learning to restore intricate details in IR images through non-explicit models is garnering increasing interest. On the one hand, Su{\'a}rez \etal~\cite{suarez2024enhancement} initially proposed the integration of sparse edge information from visible light images, along with the fusion of interpretable sparse priors, to aid in IR image reconstruction. Experimental results indicate that enhanced neural networks employing multiple receptive fields contribute to performance improvement~\cite{10379652}. On the other hand, researchers have explored the rich detail present in visible light images to improve IR image quality. This is achieved by designing modules for high-frequency information extraction from visible light images and employing attention mechanisms to judiciously introduce pattern information into the IR feature domain~\cite{liu2024dsfusion,jiang2023enhanced}. Although these methods typically follow a pre-train and fine-tune paradigm, they risk transferring undesirable features like noise from the visible light domain\cite{jiang2022multilevel,jiang2022deep}. Our study counters this by introducing target-oriented deep learning frameworks, specifically designed to overcome these limitations and enhance IR images.

\subsection{\prrr{Texture-noise Oriented} Domain Adaptation}

To enhance the restoration of intricate textures in SR tasks IR images, there is a growing trend to incorporate pattern information from visible light images. This approach has prompted an exploration into the domain adaptation challenges between the visible and IR feature domains. \prr{In this study, we proposed to treat the beneficial detailed texture information and the negative noise artifacts independently~\cite{chen2023model,shi2023source}. These two types of patterns are targeted and require respectively oriented novel approaches to further improve the SR image quality.}

Previous research~\cite{sun2016return,wei2021toalign} has shown that domain adaptation strategies can affect final performance due to the inclusion of irrelevant features from the source domain. Thus, it is crucial to explicitly decompose source domain features into task-relevant components, allowing the model to concentrate on pertinent priors~\cite{wei2021toalign,peng2019domain}. Recent advances in infrared image super-resolution have focused heavily on domain adaptation and noise handling. \hyspr{For example, Jiang \etal. introduced a framework that incorporates adaptive priors to enhance denoising performance by learning task-specific noise characteristics, which allows for the effective removal of noise without over-smoothing details. Similarly, techniques like few-shot learning for IR image super-resolution demonstrate robust performance even with minimal training data by leveraging multiscale feature recursion. These works underscore the importance of incorporating adaptive mechanisms to handle diverse artifacts, especially in IR images, where texture and noise adaptation are crucial due to low contrast and high noise levels.}

\hyspr{IR images present unique challenges that make texture and noise adaptation particularly critical. Unlike visible light images, IR images are often characterized by lower contrast, increased noise, and a lack of fine texture details due to the nature of infrared sensors, which operate in low-light or thermal environments. In this work, we introduce a dual-faceted framework aimed at facilitating feature domain alignment. Specifically, we propose texture-oriented and noise-oriented strategies. The texture-oriented approach focuses on aligning relevant texture information between the source and target domains, while the noise-oriented strategy works to reduce the transfer of undesirable noise features. This comprehensive approach aims to tackle the complexities associated with domain adaptation for IR image SR tasks.}

\section{Methodology\label{sec.3}}

\hhys{In the realm of SISR for IR imagery, traditional techniques~\cite{wang2020deep} often rely on paired training data denoted by $\left\{x_{i r}, y_{i r}\right\}$. Here, $x_{i r} \in \mathbb{R}^{\frac{H}{\mu} \times \frac{W}{\mu} \times 1}$ and $y_{i r} \in \mathbb{R}^{H \times W \times 1}$ correspond to low- and high-resolution IR images, respectively, where $\mu$ signifies the spatial upscaling factor affecting both dimensions, height $(H)$ and width $(W)$. Conversely, state-of-the-art approaches~\cite{huang2021infrared,honda2019multi,gupta2021toward} utilize a visible light counterpart, $x_{v i s} \in \mathbb{R}^{\frac{H}{\mu} \times \frac{W}{\mu} \times C}$, in a triplet formation $\left\{x_{i r}, x_{v i s}, y_{i r}\right\}$ to augment the super-resolution reconstruction of IR images. Here, $C$ signifies the RGB image's channel count, usually three.}

We now introduce the notational conventions and specialized formulations incorporated in the DASRGAN framework, engineered to tackle the SISR challenge, which is intrinsically an ill-posed problem~\cite{wang2020deep,jiang2023graph}. The analysis begins with the consideration of IR-visible pairs $\left\{x_{i r}, y_{i r}\right\}$, with $x$ denoting the LR and $y$ the HR samples. In terms of domain adaptation, the dataset is designated as $S_{v i s}:=\left\{\left(x_{v i s}, y_{i r}\right)\right\}_{i=1}^N$, where $x_{v i s}$ originating from the visible light LR distribution $\mathcal{D}_{x_{v i s}}$. Furthermore, we construct a parametric model predicated on the source dataset $S_{i r}:=\left\{\left(x_{i r}, y_{i r}\right)\right\}_{i=1}^N$, $x_{i r}$ sampled from the IR-LR distribution $\mathcal{D}_{x_{i r}}$. Note that the deployed model operates in a distinct test or target distribution, $\mathcal{D}_{x_{i r}}$, potentially diverging significantly from the source distribution, $\mathcal{D}_{x_{v i s}}$. This variation necessitates robust domain adaptation techniques, which are addressed by the target-centric DASRGAN methodology. 

\prr{In this study, we aim to enrich the texture of infrared images by integrating detailed features from visible images, simultaneously minimizing noise and artifact transmission. To achieve this, \textbf{\textit{TOA}} and \textbf{\textit{NOA}} are proposed: \textbf{\textit{1) TOA}}: We introduce a specialized discriminator, $\mathbb{D}_{s_{\text {trans }}}$, featuring a Sobel-prior extraction branch to enhance texture detail focus during adversarial training. A novel texture-oriented prior adversarial loss, $\mathcal{L}_{\text {trans }}$, is proposed to steer the generator network $\mathbb{G}$ towards precise texture pattern alignment.
\textbf{\textit{2) NOA}}: We present a new noise-oriented adversarial loss, $\mathcal{L}_n^{\mathbb{G}}$, designed to mitigate noise transfer between visible and IR domains by guiding the generator network.}

The subsequent subsections outline DASRGAN's two-stage alternating training paradigm. Sec.\ref{sec.Network Architecture} details the network architecture, while Sec.\ref{sec.Texture-Oriented Adaptation} and Sec.\ref{sec.Noise-Oriented Adaptation} elaborate on TOA and NOA, respectively.

\subsection{Training paradigm of DASRGAN}

\begin{figure}[htbp]
    \centering
      \resizebox{0.8\textwidth}{!}{%
    \includegraphics{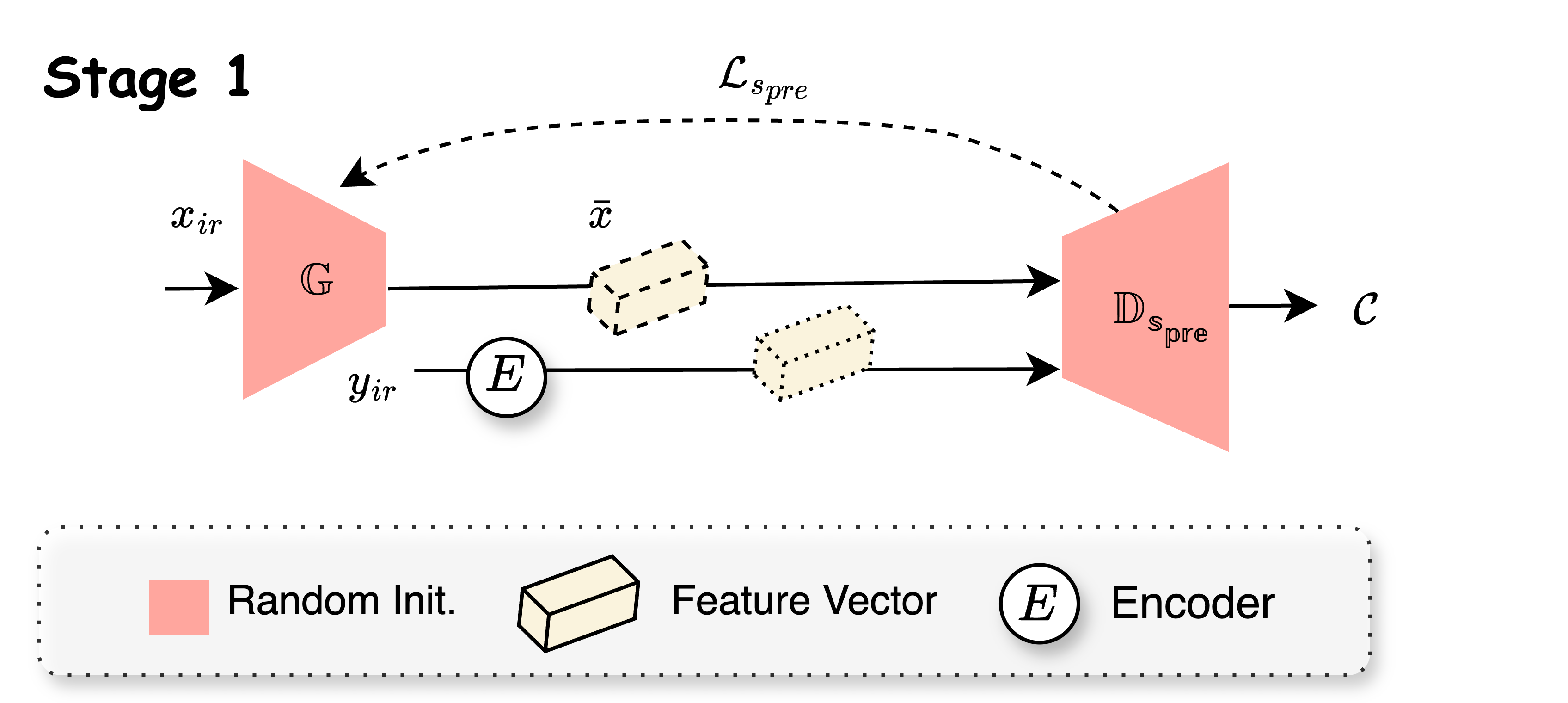}}
    \caption{\prrr{Stage 1 of DASRGAN uses adversarial learning to represent the nonlinear relationships in the paired images, where the inputs are LR-HR pairs of $\left\{x_{i r}, y_{i r}\right\}$. The figure illustrates that Basic Block (see Fig.\ref{fig rrdb}) are encoders $E$. Cross-entropy loss denote as $\mathcal{L}_{\text {spre}}$.} Best viewed in color.}
    \label{fig_method_s1}\vspace{-0.5cm}
\end{figure}

\begin{figure}[htbp]
    \centering
      \resizebox{\textwidth}{!}{%
    \includegraphics{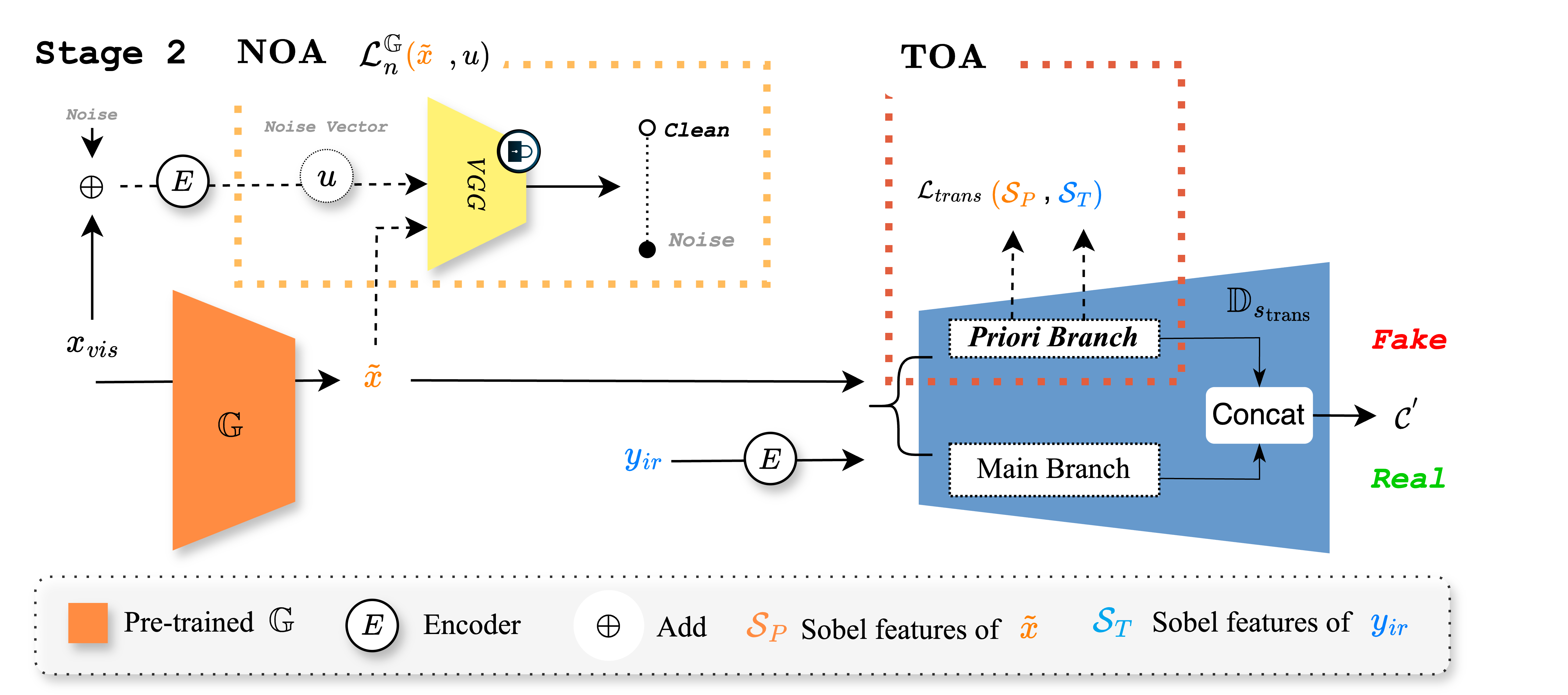}}
    \caption{\prrr{Stage 2 of DASRGAN advances with domain adaptation to refine the detail transfer from visible images via NOA and TOA. First, for NOA, the feature vector $u$ from the noise image and the output $\tilde{x}$ from the pre-trained generator $\mathbb{G}$ is fed to the pre-trained feature extractor VGG. The mid-feature layer of $u$ and $\tilde{x}$ is computed as loss in the noise adversarial loss function $\mathcal{L}_n^{\mathbb{G}}$. Second, for TOA, the novel Priori Branch is proposed to extract the Sobel features of $\tilde{x}$ and ground truth $y_{ir}$ separately. Further, their output is fed to the prior adversarial loss $\mathcal{L}_{\text {trans}}$ to encourage the model to transfer more textures.}}
    \label{fig_method_s2}\vspace{-0.5cm}
\end{figure}

DASRGAN training incorporates two round-based stages alternating IR and visible light domains. A comprehensive overview of our approach is illustrated in Fig.\ref{fig_method_s1} and Fig.\ref{fig_method_s2}. \prrr{A more detailed exposition will follow:}


\textbf{Stage 1.} \prr{At this stage, the IR images are fed to the neural network designed to capture the base IR image LR- HR mapping relationship.} The low resolution IR image $x_{i r}$ is fed to the generator network $\mathbb{G}$ to invert the input into a latent vector $\bar{x}$. Then the $\bar{x}$ and the paired high-resolution IR image $y_{i r}$ are fed to the discriminator network $\mathbb{D}_{\text {spre }}$ to output a scalar $\mathcal{C}$. \hhhys{The objective function is denoted as:}

\begin{equation}
\scalebox{0.9}{%
$\min _{\mathbb{G}} \mathcal{L}_{\mathrm{MAE}}\left(x_{i r}, y_{\mathrm{ir}}\right); ~\min_{\mathbb{D}_{\text{spre}}}\mathcal{L}_{\text{spre}} (x_{ir},y_{\text{ir}})$}
\end{equation}
where the pre-training function $\mathcal{L}_{\text {spre}}$ is the cross-entropy loss, which aims to maximize the likelihood data distribution $\mathcal{D}_{x_{i r}}$ of the paired samples in $S_{i r}$. \hhhys{$\mathbb{D}_{\text {spre }}$ is trained to minimize $\mathcal{L}_{\text {spre}}$ in an adversarial manner. Note that $\mathbb{G}$ is also trained to minimize MAE loss.}

\textbf{Stage 2.} \prrr{At this stage (see Fig.\ref{fig_method_s2}), the visible images $x_{v i s}$ are fed to a pre-trained generator network $\mathbb{G}$ aimed at introducing detailed textures in the visible images and suppressing negative factors such as noise and artifacts.} Specifically, $x_{v i s}$ are fed into $\mathbb{G}$ encoded to output latent vectors $\tilde{x} \sim \mathcal{D}_{\text {trans }}(\bar{x} \mid u, w)$. The paired data, $x_{v i s}$ and $y_{i r}$, are fed to the discriminator network $\mathbb{D}_{\mathbf{s}_{\text {trans }}}$ output scalar $\mathcal{C}^{'}$. $\mathcal{D}_{\text {trans}}$ is defined as the conditional probability distribution, \ie, the conditional probability to sample $\bar{x} $ under the prior $w$ and noise $u$ conditions. \prrr{$\mathbb{E}$ represents the operation of taking the average for all samples in the mini-batch.} Denote $\mathcal{H}$ as the hypothesis sets. We define the error of network $h \in \mathcal{H}$ on $\mathcal{D}_{\text {trans }}$ as:
\begin{equation}
\epsilon_{\mathcal{D}_{\text {trans }}}(h)=\mathbb{E}_{(x_{vis}, y_{ir}) \sim \mathcal{D}_{x_{vis}}}[\ell(h(x_{vis}), y_{ir})] 
\end{equation}

For $\ell(h(x_{vis}), y_{ir})$, where $\ell$ is the usual indicator function, we use $\mathcal{L}^{\mathbb{G}}$ to denote it is the $\ell$. Details will be discussed in Sec.\ref{sec.Experiments}.
In the domain adaptation phase, we seek to minimize the risk of error $\operatorname{min} \operatorname{Err}_{\mathcal{D}_{x_{vis}}}$, by bridging the distance between $w$ and $\tilde{x}$. When the network converges, $\mathbb{G}$ represents a nonlinear mapping between $x_{v i s}$ and $w$. The formalization is as follows:
\begin{equation}
\left.\operatorname{min} \operatorname{Err}_{\mathcal{D}_{x_{vis}}}\left(h\left(x_{vis}\right)=w\right)\right)
\end{equation}

On the other hand, we seek to learn the optimal parameters $\theta_{\mathbb{D}_{s_{\text {trans }}}}$ of the $\mathbb{D}_{s_{\text {trans }}}$ by maximizing the expected risk, $\bar{\theta}$ and $\theta^{*}$ denote the parameters from the stage 1 initialization and the parameters being updated, respectively:
\begin{equation}
\theta_{\mathbb{D}_{s_{\text {trans }}}}=\underset{\bar{\theta}, \theta^*}{\arg \max } \mathbb{E}_{\mathcal{D}_{x_{vis}}} \mathcal{L}_{\text {trans }}(x_{vis},y_{ir})
\end{equation}

\subsection{Network Architecture}
\label{sec.Network Architecture}
\begin{figure*}[t]
    \centering
    \resizebox{\textwidth}{!}{%
    \includegraphics{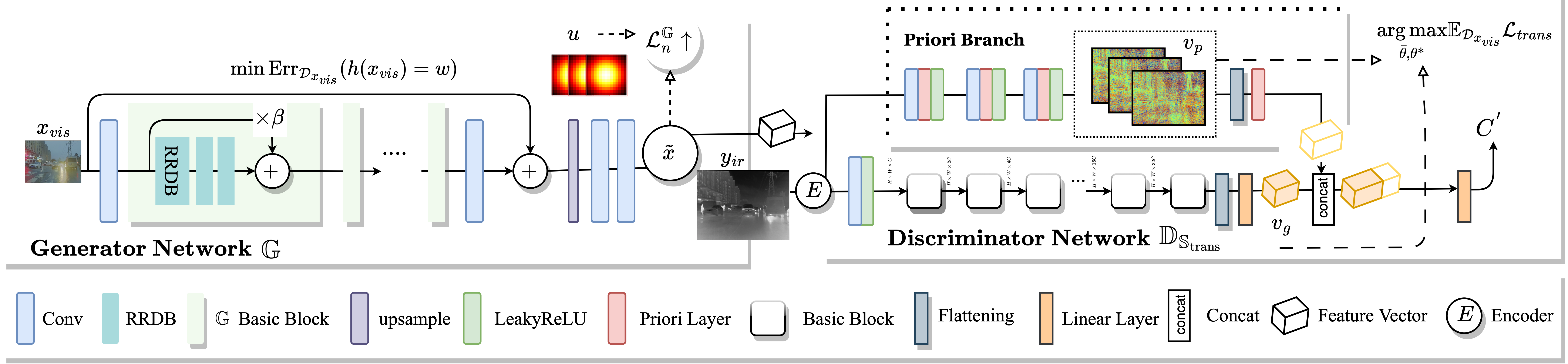}
    }
    \caption{\prrr{\textbf{Network architecture for texture-noise oriented domain adaptation adversarial training: stage 2.}} (1) For the pre-trained $\mathbb{G}$, the visible images $x_{v i s}$ are fed and then encoded into $\tilde{x}$. \prr{$x_{v i s}$ introduces Gaussian noise to the image before processing it through VGG, where the feature vector captured in the intermediate feature layer is identified as $u$.} The loss function $\mathcal{L}_n^{\mathbb{G}}$ is designed to evaluate the difference between the degradation information in $u$ and $\tilde{x}$, and to minimize $u$ through optimization. (2) For $\mathbb{D}_{\mathbf{S}_{\text {trans }}}$, novel prior branch is proposed. \pr{$\tilde{x}$ is encoded by the prior branch and outputs a latent vector $v_p$, which includes information about the details of the visible domain, \eg, edges. \hyspr{This branch is composed of a sequence of blocks, each containing a $3 \times 3$ convolutional layer, a Sobel filter, and LeakyReLU activation.} After concatenating with the main path feature representation $v_g$, $v_p$ is converted to new feature vectors and mapped to a scalar $\mathcal{C}^{'} $.}}
    \label{fig Stage 2}
\end{figure*}

In this study, our primary objective is not to develop complex generator architectures for enhanced performance in IR super-resolution tasks. Instead, we introduce a novel discriminator, $\mathbb{D}{s{\text {trans }}}$, during the domain adaptation phase. This discriminator is specifically designed to facilitate \prrr{texture-noise oriented} domain adaptation, with a special emphasis on enhancing texture details. An illustration of network architecture for texture-noise oriented domain adaptation adversarial training can be found in Fig.\ref{fig Stage 2}.

\begin{figure}[htbp]
    \centering
    \resizebox{\textwidth}{!}{%
    \includegraphics{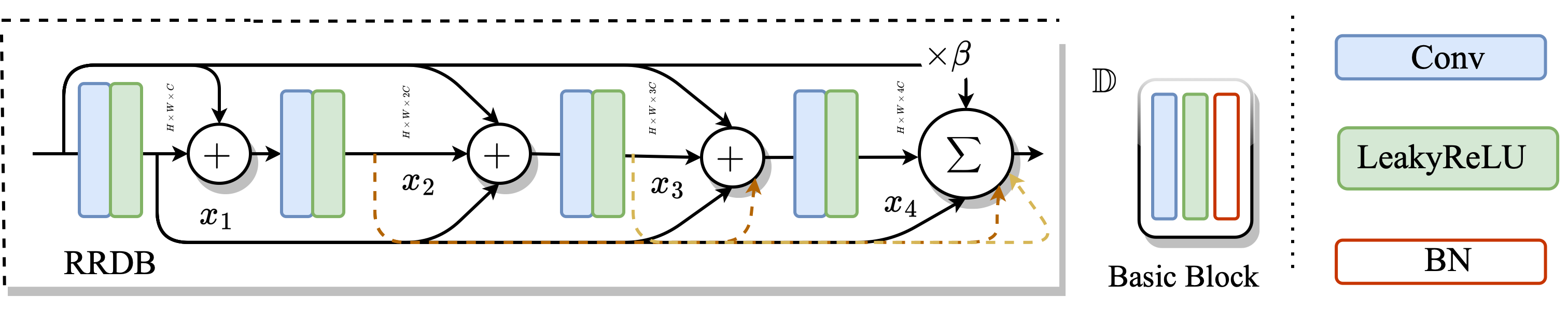}
    }
    \caption{The Architecture of Residual in Residual Dense Block (RRDB) and Basic Block.}
    \label{fig rrdb}
\end{figure}

\hyspr{The generator $\mathbb{G}$ utilizes a network architecture centered around the Residual-in-Residual Dense Block (RRDB) as the core module, as established in previous works~\cite{wang2018esrgan,wang2021real}. The RRDB block, depicted in Fig.\ref{fig rrdb}, integrates dense connections with residual scaling and omits batch normalization to enhance image quality. The generator employs 23 RRDB blocks, each leveraging skip connections and sub-pixel convolution layers for effective up-sampling. This architecture has been proven effective in fitting super-resolution datasets, particularly in both Gaussian-based and blind SR tasks~\cite{wang2021real}.}

Regarding the discriminator, we present a novel $\mathbb{D}_{s{\text {trans }}}$, specifically crafted for the domain adaptation phase. This design includes an a priori branch $\mathcal{B}$, engineered to encode the detailed texture of $x_{vis}$ into an a priori latent variable $v_{p}$. Employing the Sobel operator~\cite{sun2024unsupervised,jing2024boosting} as the a priori layer, it excels in capturing texture detail patterns, given the notable efficiency of the Sobel operator in edge detection~\cite{sun2024unsupervised,kanopoulos1988design} and detail texturing. \hyspr{Sobel priori branch is structured around a block sequence that includes a $3 \times 3$ convolutional layer, a Sobel filter layer, and LeakyReLU activation.} Our ablation studies suggest that extracting mid-level textures is most effective, which will be discussed further in Sec.\ref{sec.Experiments}. The architecture of $\mathbb{D}_{s{\text {trans }}}$ can be mathematically formalized as follows:
\vspace{-0.2cm}
\begin{equation}
C^{\prime}=L\left(\operatorname{concat}\left(\mathcal{B}\left(x_{v i s}\right), \mathcal{M}\left(x_{v i s}\right)\right)\right)
\end{equation}
where $\mathcal{M}$ serves as the primary feature extraction branch. After encoding, latent variables are concatenated $\text { concat }$ and fed into a linear layer $L$ for feature mapping. $\mathcal{C}^{'} $ is the scalar output produced by the discriminator. Details on the texture-oriented and noise-oriented domain adaptation losses, integral to our framework, will be elaborated in Sec.\ref{sec.Texture-Oriented Adaptation} and Sec.\ref{sec.Noise-Oriented Adaptation}, respectively.

\subsection{Texture-Oriented Adaptation}
\label{sec.Texture-Oriented Adaptation}

For TOA, our objective is to reconcile the distributional disparities between visible and IR images with respect to texture priors. $\mathbb{D}_{s_{\text {trans }}}$ is designed to guide the generator $\mathbb{G}$ in synthesizing realistic texture priors. This is achieved by maximizing the expected risk $\text { Err}$ through adversarial training, formulated as follows:

\begin{equation}
\scalebox{0.98}{%
$\begin{aligned}
\text{Err}\left[\mathcal{D}_{x_{ir}}, \theta, \ell(x_{ir}, y_{ir}, \theta)\right] &= \mathbb{E}_{(x_{ir}, y_{ir}) \sim \mathcal{D}_{x_{ir}}}[\ell(x_{ir}, y_{ir}, \theta)] \\
&= \mathbb{E}_{(x_{vis}, y_{ir}) \sim \mathcal{D}_{x_{vis}}}\left[\frac{\mathcal{D}_{x_{ir}}}{\mathcal{D}_{x_{vis}}} \ell \right] \\
&= \mathbb{E}_{(x_{vis}, y_{ir}) \sim \mathcal{D}_{x_{vis}}}[\beta \ell]
\end{aligned}$%
}
\end{equation}

Here, $\mathcal{D}_{x_{vis}}$ and $\mathcal{D}_{x_{ir}}$ denote the probability distributions of $S_{vis}$ and $S_{i r}$, respectively. $\ell$ represents the loss function, and $\beta(x_{vis}, x_{ir})$ is the ratio between the two probability distributions, effectively serving as a weighting coefficient. Notably, when $\mathcal{D}_{x_{vis}}=\mathcal{D}_{x_{ir}}$, $ \beta$ simplifies to 1. Given that the dense estimate $\beta$ remains constant, the texture-oriented loss function $\ell$ is expected to enhance the performance of domain adaptation. In our work, we introduce the prior adversarial loss $\mathcal{L}_{\text {trans}}$ as $\ell$ to evaluate the quality of predicted images in prior-detection. Specifically, we employ the Sobel operator to extract edge features from both predicted and ground-truth images $I$. The Sobel features, denoted as $\mathcal{S}_P$ and $\mathcal{S}_T$ for the feature vectors in predicted and target images respectively, are computed using the formula:

\begin{equation}
\mathcal{S}=\sqrt{\left(G_h \odot  I\right)^2+\left(G_v \odot  I\right)^2}
\end{equation}
where $G_h$ and $G_v$ are the Sobel kernels for horizontal and vertical edge detection, and $\odot $ represents the convolution operation. The $\mathcal{L}_{\text {trans }}$ between these Sobel features is then calculated and normalized by the total number of pixels $N$:

\begin{equation}
\mathcal{L}_{\text {trans }}=\frac{1}{N} \sum_{i, j}\left|\mathcal{S}_P(i, j)-\mathcal{S}_T(i, j)\right|
\end{equation}

\subsection{Noise-Oriented Adaptation}
\label{sec.Noise-Oriented Adaptation}

For NOA, our approach aims to address the issue of blurring degradation when transferring features from the visible light domain to the IR domain. 
To achieve this, we incorporate a noise adversarial loss function $\mathcal{L}_n^{\mathbb{G}}$ designed to quantify the perceptual divergence between $\tilde{x}$ and an artificially generated noise pattern from $u$. Specifically, we utilize a pre-trained VGG~\cite{sengupta2019going} network to extract hierarchical features, denoted as $f_k(\tilde{x})$ and $f_k (u)$, from both the predicted image and the noise pattern at the $k^{th}$ layer. The loss in each layer $k$ is calculated as the negative $L_{1}$ distance between these feature vectors, scaled by a layer-specific weight $w_k$. Formally, the $\mathcal{L}_n^{\mathbb{G}}$ is defined as:
\vspace{-0.2cm}
\begin{equation}
\mathcal{L}_n^{\mathbb{G}}(\tilde{x}, u)=-\sum_k w_k \times \| f_k(\tilde{x})-f_k(u) \|_1
\end{equation}


\hyspr{The overall loss function is a weighted combination of the noise adversarial loss and the texture-oriented adversarial loss. After conducting extensive ablation studies, we empirically set the weight for the noise adversarial loss to 0.1 and the texture-oriented adversarial loss to 1.0. This configuration was chosen as it provided the best balance between noise suppression and texture preservation.} This specialized loss function serves to guide the network toward generating images that are perceptually different from natural noise patterns, thereby enhancing the model's adaptation ability. This is particularly advantageous in scenarios involving Gaussian blurred scenes, a common occurrence in the visible feature domain. The objective function of DASRGAN at stage 2 is

\begin{equation}
\scalebox{0.8}{%
$\begin{aligned}
\min _\mathbb{G} \max _{\mathbb{D}_{s_{\text {trans }}}} V(\mathbb{D}_{s_{\text {trans }}}, \mathbb{G})= & E_{y_{ir}}[\log \mathbb{D}_{s_{\text {trans }}}(y_{ir})]  +E_{x_{vis}}[\log (1-\mathbb{D}_{s_{\text {trans }}}(\mathbb{G}(x_{vis})))]
\end{aligned}$
}
\end{equation}
where $\log \mathbb{D}_{s_{\text {trans }}}(y_{ir})$ is the cross-entropy between $\left[1,0\right]^T$ and $[\mathbb{D}_{s_{\text {trans }}}\left(y_{ir}\right), 1-\mathbb{D}_{s_{\text {trans }}}\left(y_{ir}\right)]^T$. Similarly, $\log (1-\mathbb{D}_{s_{\text {trans }}}(\mathbb{G}(y_{i r})))$ is the cross-entropy between $\left[1,0\right]^T$ and $[\left.\mathbb{D}_{s_{\text {trans }}}\left(\mathbb{G}\left(x_{vis}\right)\right)\right), 1-\mathbb{D}_{s_{\text {trans }}}(\mathbb{G}(x_{vis})))]^T$.

\section{Experiments}
\label{sec.Experiments}

\subsection{\pr{Experimental Settings}}

\begin{figure}[htbp]
    \centering
    \resizebox{\textwidth}{!}{%
    \includegraphics{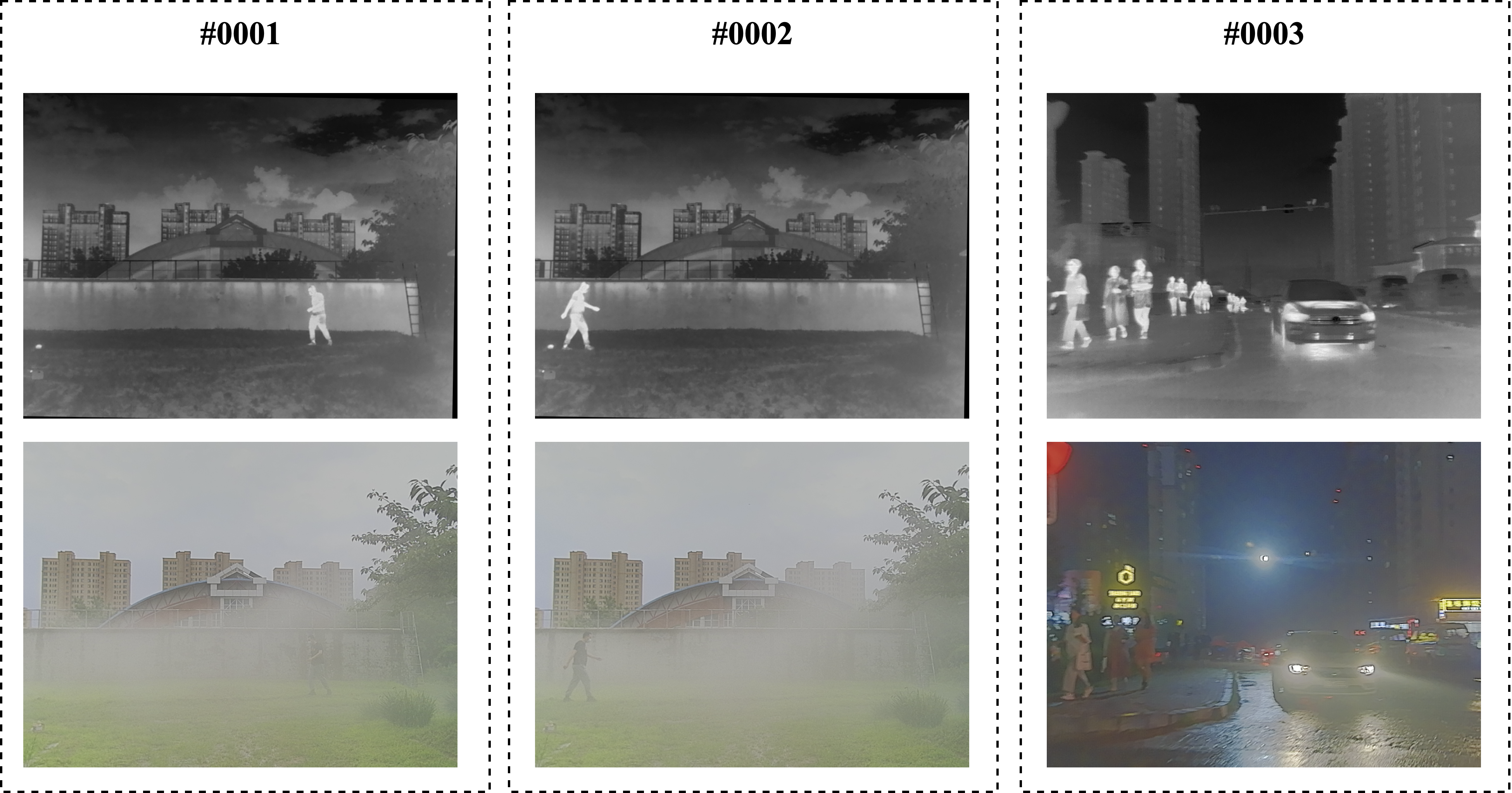}
    }
    \caption{Visualization of infrared-visible images on the training dataset.}
    \label{dataset fig}
\end{figure}

\textbf{Datasets.} In the training phase, we curated a paired dataset derived from Liu \etal~\cite{liu2022target}: M3FD, intricately compiled with a synchronized imaging system equipped with well-calibrated infrared and optical sensors, includes 4,177 aligned infrared-visible image pairs and 23,635 annotated objects. The dataset's diverse range covers various environments, illumination conditions, seasons, and weather scenarios, providing a rich pixel variation as depicted in Fig.\ref{dataset fig} of our study. For evaluation, we created three specialized datasets: M3FD5, M3FD15, and M3FD20, from the M3FD dataset, each with different sample sizes and scene compositions. Additionally, to examine the model's generalization performance, we compiled two more test datasets, CVC5 and CVC15, from the CVC dataset~\cite{campo2012multimodal}. \textbf{Metrics.} \hyspr{Consistent with prior research, both the training and test datasets were generated using bicubic downsampling to produce LR paired samples. The evaluation phase involved assessing the generated samples using key performance metrics such as Peak Signal-to-Noise Ratio (PSNR), Mean Squared Error (MSE), and Structural Similarity Index Measure (SSIM).} 

\hyspr{PSNR was calculated as:}

\begin{equation}
\operatorname{PSNR}=10 \cdot \log _{10}\left(\frac{\mathrm{MAX}^2}{\mathrm{MSE}}\right)
\end{equation} \hyspr{where MAX represents the maximum possible pixel value, and MSE is the Mean Squared Error between the ground truth (HR) image and the SR image. MSE was computed by averaging the squared differences between the HR image and the SR image:}

\begin{equation}
\operatorname{MSE}=\frac{1}{H \times W} \sum_{i=1}^H \sum_{j=1}^W\left(I_{\mathrm{HR}}(i, j)-I_{\mathrm{SR}}(i, j)\right)^2
\end{equation} \hyspr{where $H$ and $W$ denote the height and width of the image, and $I_{\mathrm{HR}}$ and $I_{\mathrm{SR}}$ are the pixel values of the HR and SR images, respectively. SSIM was used to assess the perceptual quality by comparing luminance, contrast, and structure between the HR and SR images:}

\begin{equation}
\operatorname{SSIM}\left(I_{\mathrm{HR}}, I_{\mathrm{SR}}\right)=\frac{\left(2 \mu_{\mathrm{HR}} \mu_{\mathrm{SR}}+C_1\right)\left(2 \sigma_{\mathrm{HR}, \mathrm{SR}}+C_2\right)}{\left(\mu_{\mathrm{HR}}^2+\mu_{\mathrm{SR}}^2+C_1\right)\left(\sigma_{\mathrm{HR}}^2+\sigma_{\mathrm{SR}}^2+C_2\right)}
\end{equation} \hyspr{where $\mu_{\mathrm{HR}}$ and $\mu_{\mathrm{SR}}$ are the mean intensities, $\sigma_{\mathrm{HR}}$ and $\sigma_{\mathrm{SR}}$ are the variances, and $\sigma_{\mathrm{HR}, \mathrm{SR}}$ is the covariance of the HR and SR images. The constants $C_1$ and $C_2$ are included to stabilize the division.}

\textbf{Implementation details.}
In each minimal training batch, randomly cropped $64 \times 64$ LR images were used as input and optimized using Adam. The learning rate was set to 1e-5. Experiments were developed in the PyTorch framework on an Nvidia A6000 GPU, based on the BasicSR toolbox.

\begin{table*}[t]
\centering
\caption{\hyspr{Quantitative comparison (PSNR$\uparrow$ MSE$\downarrow$ SSIM$\uparrow$). For Single image super-resolution on M3FD datasets. Best and second-best performances are marked in \textbf{bold} and {\ul underlined}, respectively. The bottom method marked in \textcolor{gray}{gray} adopts our model.}}
\resizebox{\textwidth}{!}{%
\begin{tabular}{@{}c|lccccccccccc@{}}
\toprule
                                           & \multicolumn{1}{c|}{}                          & \multicolumn{2}{c|}{}                           & \multicolumn{3}{c|}{M3FD5}                                                                                                                                                      & \multicolumn{3}{c|}{M3FD15}                                                                                                                                                              & \multicolumn{3}{c}{M3FD20}                                                                                                                                                               \\ \cmidrule(l){5-13} 
\multirow{-2}{*}{Scale}                    & \multicolumn{1}{c|}{\multirow{-2}{*}{Methods}} & \multicolumn{1}{c}{\multirow{-2}{*}{\#Params[K]}} & \multicolumn{1}{c|}{\multirow{-2}{*}{\#FLOPs[G]}} & PSNR$\uparrow $                                                    & MSE$\downarrow $                                                      & \multicolumn{1}{c|}{SSIM$\uparrow $}                                & PSNR$\uparrow $                                                          & MSE$\downarrow $                                                         & \multicolumn{1}{c|}{SSIM$\uparrow $}                                   & PSNR$\uparrow $                                                          & MSE$\downarrow$                                                        & SSIM$\uparrow $                                                        \\ \midrule
                                           & EDSR\textcolor[RGB]{217,205,144}{\textit{[CVPRW 2017]}}~\cite{lim2017enhanced}                                            & 1,369                                          & 316                                         & 46.8033                                                   & 1.8877                                                   & 0.9903                                                   & 47.5681                                                      & 1.1512                                                      & 0.9906                                                      & 47.6258                                                      & 0.9873                                                      & 0.9915                                                      \\
                                           & ESRGAN\textcolor[RGB]{217,205,144}{\textit{[ECCVW 2018]}}~\cite{wang2018esrgan}                                          & 16,661                                         & 207                                        & 47.7307                                                   & 1.3792                                                   & 0.9920                                                   & 48.0642                                                      & 1.0141                                                      & 0.9910                                                      & 48.0353                                                      & 0.8923                                                      & 0.9918                                                      \\
                                           & FSRCNN\textcolor[RGB]{217,205,144}{\textit{[ECCV 2016]}}~\cite{dong2016accelerating}                                         & 475                                            & 6                                           & 45.4145                                                   & 2.5077                                                   & 0.9892                                                   & 45.3984                                                      & 1.8363                                                      & 0.9901                                                      & 46.0699                                                      & 1.4521                                                      & 0.9911                                                      \\
                                           & SRGAN\textcolor[RGB]{217,205,144}{\textit{[CVPR 2017]}}~\cite{ledig2017photo}                                          & 1,369                                          & 220                                          & 46.7851                                                   & 1.8836                                                   & 0.9903                                                   & 47.7077                                                      & 1.1141                                                      & 0.9906                                                      & 47.5877                                                      & 0.9867                                                      & 0.9915                                                      \\
                                           & SwinIR\textcolor[RGB]{217,205,144}{\textit{[ICCV 2021]}}~\cite{liang2021swinir}                                         & 11,752                                         & 49                                        & 46.7244                                                   & 1.8973                                                   & 0.9902                                                   & 47.1768                                                      & 1.2517                                                      & 0.9901                                                      & 47.4024                                                      & 1.0424                                                      & 0.9912                                                      \\
                                           & SRCNN\textcolor[RGB]{217,205,144}{\textit{[T-PAMI 2015]}}~\cite{dong2015image}                                          & 57                                             & 53                                             & 45.2227                                                   & 2.6043                                                   & 0.9889                                                   & 45.2058                                                      & 1.9163                                                      & 0.9899                                                      & 46.0374                                                      & 1.4761                                                      & 0.9910                                                      \\
                                           & RCAN\textcolor[RGB]{217,205,144}{\textit{[ECCV 2018]}}~\cite{zhang2018image}                                            & 12,467                                         & 38                                        & 46.8454                                                   & 1.7348                                                   & 0.9911                                                   & 47.5058                                                      & 1.1610                                                      & 0.9909                                                      & 47.5261                                                      & 1.0085                                                      & 0.9918                                                      \\
                                           & PSRGAN\textcolor[RGB]{217,205,144}{\textit{[SPL 2021]}}~\cite{huang2021infrared}                                          & 2,414                                          & 218                                          & 46.3526                                                   & 2.2219                                                   & 0.9895                                                   & 47.3337                                                      & 1.2204                                                      & 0.9903                                                      & 47.2350                                                      & 1.0842                                                      & 0.9912                                                      \\
                                           & Shuffle (base)\textcolor[RGB]{217,205,144}{\textit{[NIPS'22]}}~\cite{sun2022shufflemixer}                            & 393                                            & 43                                            & 47.2429                                                   & 1.5639                                                   & 0.9913                                                   & 47.7097                                                      & 1.1100                                                      & 0.9907                                                      & 47.7688                                                      & 0.9554                                                      & 0.9915                                                      \\
                                           & Shuffle (tiny)\textcolor[RGB]{217,205,144}{\textit{[NIPS'22]}}~\cite{sun2022shufflemixer}                            & 108                                            & 12                                           & 46.7761                                                   & 1.8953                                                   & 0.9902                                                   & 47.5158                                                      & 1.1563                                                      & 0.9906                                                      & 47.5547                                                      & 0.9959                                                      & 0.9914                                                      \\
                                           & SAFMN\textcolor[RGB]{217,205,144}{\textit{[ICCV 2023]}}~\cite{sun2023spatially}                                     & 5,559                                          & 52                                         & 48.0102                                                   & 1.0881                                                   & 0.9917                                                   & 48.3006                                                      & 0.8202                                                      & 0.9909                                                      & 48.4714                                                      & 0.6981                                                      & 0.9917                                                      \\
                                           & HAT \textcolor[RGB]{217,205,144}{\textit{[CVPR 2023]}}~\cite{chen2023activating}                                       & 20,624                                         & 103                                         & {\ul 48.3822}                                             &  \textbf{0.9813}                                             & \textbf{0.9923}                                          & {\ul 48.4636}                                                & {\ul 0.7720}                                                & {\ul 0.9914}                                                & {\ul 48.7168}                                                & {\ul 0.6545}                                                & \textbf{0.9920}                                             \\
\multirow{-13}{*}{$\times 2$}                      & \cellcolor[HTML]{EFEFEF}Ours                   & \cellcolor[HTML]{EFEFEF}16,661                 & \cellcolor[HTML]{EFEFEF}207                 & \cellcolor[HTML]{EFEFEF}\textbf{48.3865}                  & \cellcolor[HTML]{EFEFEF}{\ul0.9982}                  & \cellcolor[HTML]{EFEFEF}{\ul 0.9922}                     & \cellcolor[HTML]{EFEFEF}\textbf{48.8402}                     & \cellcolor[HTML]{EFEFEF}\textbf{0.7148}                     & \cellcolor[HTML]{EFEFEF}\textbf{0.9915}                     & \cellcolor[HTML]{EFEFEF}\textbf{48.7952}                     & \cellcolor[HTML]{EFEFEF}\textbf{0.6410}                     & \cellcolor[HTML]{EFEFEF}{\ul 0.9919}                        \\ \midrule
\multicolumn{1}{l|}{}                      & EDSR\textcolor[RGB]{217,205,144}{\textit{[CVPRW 2017]}}~\cite{lim2017enhanced}                                            & 1,517                                          & 316                                          & \multicolumn{1}{l}{39.1096}                               & \multicolumn{1}{l}{13.6760}                              & \multicolumn{1}{l}{0.9523}                               & \multicolumn{1}{l}{40.6962}                                  & \multicolumn{1}{l}{5.4159}                                  & \multicolumn{1}{l}{0.9712}                                  & \multicolumn{1}{l}{40.9810}                                  & \multicolumn{1}{l}{4.8430}                                  & \multicolumn{1}{l}{0.9742}                                  \\
\multicolumn{1}{l|}{}                      & ESRGAN\textcolor[RGB]{217,205,144}{\textit{[ECCVW 2018]}}~\cite{wang2018esrgan}                                         & 16,697                                         & 207                                      & \multicolumn{1}{l}{40.2261}                               & \multicolumn{1}{l}{9.3695}                               & \multicolumn{1}{l}{0.9544}                               & \multicolumn{1}{l}{{\ul 41.8889}}                            & \multicolumn{1}{l}{{\ul 3.5478}}                            & \multicolumn{1}{l}{{\ul 0.9720}}                            & \multicolumn{1}{l}{{\ul 42.1464}}                            & \multicolumn{1}{l}{{\ul 3.1274}}                            & \multicolumn{1}{l}{{\ul 0.9751}}                            \\
\multicolumn{1}{l|}{}                      & FSRCNN\textcolor[RGB]{217,205,144}{\textit{[ECCV 2016]}}~\cite{dong2016accelerating}                                          & 623                                            & 6                                            & \multicolumn{1}{l}{39.1111}                               & \multicolumn{1}{l}{10.8185}                              & \multicolumn{1}{l}{0.9509}                               & \multicolumn{1}{l}{40.1994}                                  & \multicolumn{1}{l}{5.1549}                                  & \multicolumn{1}{l}{0.9703}                                  & \multicolumn{1}{l}{40.9195}                                  & \multicolumn{1}{l}{4.2574}                                  & \multicolumn{1}{l}{0.9734}                                  \\
\multicolumn{1}{l|}{}                      & SRGAN\textcolor[RGB]{217,205,144}{\textit{[CVPR 2017]}}~\cite{ledig2017photo}                                          & 1,517                                          & 220                                          & \multicolumn{1}{l}{40.0669}                               & \multicolumn{1}{l}{9.7075}                               & \multicolumn{1}{l}{0.9531}                               & \multicolumn{1}{l}{41.8255}                                  & \multicolumn{1}{l}{3.5842}                                  & \multicolumn{1}{l}{0.9717}                                  & \multicolumn{1}{l}{42.0157}                                  & \multicolumn{1}{l}{3.2603}                                  & \multicolumn{1}{l}{0.9745}                                  \\
\multicolumn{1}{l|}{}                      & SwinIR\textcolor[RGB]{217,205,144}{\textit{[ICCV 2021]}}~\cite{liang2021swinir}                                          & 11,900                                         & 49                                         & \multicolumn{1}{l}{39.1953}                               & \multicolumn{1}{l}{10.6050}                              & \multicolumn{1}{l}{0.9520}                               & \multicolumn{1}{l}{40.4065}                                  & \multicolumn{1}{l}{4.8949}                                  & \multicolumn{1}{l}{0.9710}                                  & \multicolumn{1}{l}{40.9260}                                  & \multicolumn{1}{l}{4.1474}                                  & \multicolumn{1}{l}{0.9741}                                  \\
\multicolumn{1}{l|}{}                      & SRCNN\textcolor[RGB]{217,205,144}{\textit{[T-PAMI 2015]}}~\cite{dong2015image}                                          & 168                                            & 53                                            & \multicolumn{1}{l}{37.2484}                               & \multicolumn{1}{l}{14.2045}                              & \multicolumn{1}{l}{0.9400}                               & \multicolumn{1}{l}{37.7955}                                  & \multicolumn{1}{l}{9.2622}                                  & \multicolumn{1}{l}{0.9601}                                  & \multicolumn{1}{l}{39.0011}                                  & \multicolumn{1}{l}{6.9846}                                  & \multicolumn{1}{l}{0.9652}                                  \\
\multicolumn{1}{l|}{}                      & RCAN\textcolor[RGB]{217,205,144}{\textit{[ECCV 2018]}}~\cite{zhang2018image}                                            & 12,614                                         & 38                                        & \multicolumn{1}{l}{39.3303}                               & \multicolumn{1}{l}{10.7535}                              & \multicolumn{1}{l}{0.9519}                               & \multicolumn{1}{l}{40.3756}                                  & \multicolumn{1}{l}{4.9195}                                  & \multicolumn{1}{l}{0.9713}                                  & \multicolumn{1}{l}{41.1036}                                  & \multicolumn{1}{l}{4.1006}                                  & \multicolumn{1}{l}{0.9746}                                  \\
\multicolumn{1}{l|}{}                      & PSRGAN\textcolor[RGB]{217,205,144}{\textit{[SPL 2021]}}~\cite{huang2021infrared}                                          & 2,414                                          & 207                                          & \multicolumn{1}{l}{39.4035}                               & \multicolumn{1}{l}{10.3635}                              & \multicolumn{1}{l}{0.9518}                               & \multicolumn{1}{l}{40.7592}                                  & \multicolumn{1}{l}{4.4829}                                  & \multicolumn{1}{l}{0.9709}                                  & \multicolumn{1}{l}{41.3344}                                  & \multicolumn{1}{l}{3.7767}                                  & \multicolumn{1}{l}{0.9738}                                  \\
\multicolumn{1}{l|}{}                      & Shuffle (base)\textcolor[RGB]{217,205,144}{\textit{[NIPS'22]}}~\cite{sun2022shufflemixer}                            & 410                                            & 43                                            & \multicolumn{1}{l}{40.1626}                               & \multicolumn{1}{l}{9.0440}                               & \multicolumn{1}{l}{0.9553}                               & \multicolumn{1}{l}{41.5322}                                  & \multicolumn{1}{l}{3.8124}                                  & \multicolumn{1}{l}{0.9714}                                  & \multicolumn{1}{l}{41.8758}                                  & \multicolumn{1}{l}{3.4093}                                  & \multicolumn{1}{l}{0.9747}                                  \\
\multicolumn{1}{l|}{}                      & Shuffle (tiny)\textcolor[RGB]{217,205,144}{\textit{[NIPS'22]}}~\cite{sun2022shufflemixer}                            & 112                                            & 12                                            & \multicolumn{1}{l}{39.7708}                               & \multicolumn{1}{l}{9.9019}                               & \multicolumn{1}{l}{0.9530}                               & \multicolumn{1}{l}{41.2227}                                  & \multicolumn{1}{l}{4.0582}                                  & \multicolumn{1}{l}{0.9714}                                  & \multicolumn{1}{l}{41.6794}                                  & \multicolumn{1}{l}{3.5265}                                  & \multicolumn{1}{l}{0.9743}                                  \\
\multicolumn{1}{l|}{}                      & HAT \textcolor[RGB]{217,205,144}{\textit{[CVPR 2023]}}~\cite{chen2023activating}                                       & 20,772                                         & 103                                         & \multicolumn{1}{l}{39.8073}                               & \multicolumn{1}{l}{9.8251}                               & \multicolumn{1}{l}{0.9527}                               & \multicolumn{1}{l}{41.2769}                                  & \multicolumn{1}{l}{4.0311}                                  & \multicolumn{1}{l}{0.9709}                                  & \multicolumn{1}{l}{41.7172}                                  & \multicolumn{1}{l}{3.4732}                                  & \multicolumn{1}{l}{0.9742}                                  \\
\multicolumn{1}{l|}{}                      & SAFMN\textcolor[RGB]{217,205,144}{\textit{[ICCV 2023]}}~\cite{sun2023spatially}                                     & 5,600                                          & 52                                          & \multicolumn{1}{l}{\textbf{40.5748}}                      & \multicolumn{1}{l}{\textbf{7.2981}}                      & \multicolumn{1}{l}{\textbf{0.9604}}                      & \multicolumn{1}{l}{41.3232}                                  & \multicolumn{1}{l}{4.0054}                                  & \multicolumn{1}{l}{0.9713}                                  & \multicolumn{1}{l}{41.9802}                                  & \multicolumn{1}{l}{3.3822}                                  & \multicolumn{1}{l}{0.9748}                                  \\
\multicolumn{1}{l|}{\multirow{-13}{*}{$\times 4$}} & \cellcolor[HTML]{EFEFEF}Ours                   & \cellcolor[HTML]{EFEFEF}16,697                 & \cellcolor[HTML]{EFEFEF}207                 & \multicolumn{1}{l}{\cellcolor[HTML]{EFEFEF}{\ul 40.5088}} & \multicolumn{1}{l}{\cellcolor[HTML]{EFEFEF}{\ul 8.7855}} & \multicolumn{1}{l}{\cellcolor[HTML]{EFEFEF}{\ul 0.9558}} & \multicolumn{1}{l}{\cellcolor[HTML]{EFEFEF}\textbf{42.1196}} & \multicolumn{1}{l}{\cellcolor[HTML]{EFEFEF}\textbf{3.3763}} & \multicolumn{1}{l}{\cellcolor[HTML]{EFEFEF}\textbf{0.9724}} & \multicolumn{1}{l}{\cellcolor[HTML]{EFEFEF}\textbf{42.3179}} & \multicolumn{1}{l}{\cellcolor[HTML]{EFEFEF}\textbf{3.0009}} & \multicolumn{1}{l}{\cellcolor[HTML]{EFEFEF}\textbf{0.9755}} \\ \bottomrule
\end{tabular}%
}
\label{tab:Quantitative M3FD}
\end{table*}

\textbf{Loss function.} For discriminator loss $\mathcal{L}^{\mathbb{D}}$, \hhhys{cross-entropy loss and prior adversarial loss} $\mathcal{L}_{\text {trans }}$ are used. \hhhys{Discriminator maximizes the difference between real and predicted distributions, using a negative loss setting to minimize the loss function and optimize parameters.} For the generator loss $\mathcal{L}^{\mathbb{G}}$, MAE and noise adversarial losses $\mathcal{L}_n^{\mathbb{G}}$ are used.  $\alpha$ and $\beta$ \hhhys{are the balancing weights:}

\begin{equation}
\begin{aligned}
& \mathcal{L}^{\mathbb{G}}=\mathcal{L}_{M A E}+\alpha \mathcal{L}_n^{\mathbb{G}}\\
& \mathcal{L}^{\mathbb{D}}=-(\mathcal{L}_{spre}+\beta \mathcal{L}_{\text {trans }})
\end{aligned}
\end{equation}

\subsection{Quantitative Results}

Tab.\ref{tab:Quantitative M3FD} provides a comprehensive quantitative comparison of various SISR methods on the M3FD datasets. Evaluation metrics include PSNR, MSE, and SSIM. The evaluation of various methods is performed on multiple scales ($\times 2$ and $\times 4$) and distinct versions of the M3FD dataset (M3FD5, M3FD15, M3FD20). Additionally, the number of parameters for each model is disclosed, offering a lens into the complexity of the respective models. Remarkably, our proposed method (highlighted in gray) consistently outshines other state-of-the-art techniques across nearly all performance metrics and dataset versions. It registers the highest PSNR and SSIM scores while simultaneously achieving the lowest MSE, thereby indicating exceptional image quality and structural integrity. In terms of parameter efficiency, our model demonstrates superior performance despite having a parameter count comparable to that of ESRGAN~\cite{wang2018esrgan}, thus indicating optimized parameter utilization. Contemporary methods such as HAT~\cite{chen2023activating} and SAFMN~\cite{sun2023spatially} exhibit competitive performance but ultimately fall short of the benchmarks set by our method. Legacy algorithms like SRCNN~\cite{dong2015image} and FSRCNN~\cite{dong2016accelerating} lag in performance metrics, underscoring the DASRGAN advancements in the field of IR image super-resolution. Importantly, our model maintains a consistent lead across different scales, reinforcing its robustness and scalability. 

To rigorously assess DASRGAN's robustness, we conducted tests on the CVC dataset (see Tab.\ref{tab:Quantitative CVC}). Our model consistently outperforms benchmarks on all scales in both the CVC15 and CVC5 datasets, achieving the highest PSNR, the lowest MSE, and the top-ranking SSIM. This establishes its state-of-the-art status. In terms of scale robustness, our model sustains its superior performance across varying scales ($\times 2$ \& $\times 4$), reinforcing its adaptability. While HAT~\cite{chen2023activating} and Shuffle (tiny)~\cite{sun2022shufflemixer} show competitive results, they are outperformed by our approach. In contrast, SAFMN~\cite{sun2023spatially} lags significantly, indicating limited robustness in various IR feature domains. Our model excels across all key metrics, underscoring its comprehensive effectiveness for IR image super-resolution tasks.

\begin{table}[htbp]
\centering
\caption{Quantitative comparison (PSNR$\uparrow$ MSE$\downarrow$ SSIM$\uparrow$). For IR image super-resolution on CVC datasets. Best and second-best performances are marked in \textbf{bold} and {\ul underlined}, respectively. The bottom method marked in \textcolor{gray}{gray} adopts our model.}
\label{tab:Quantitative CVC}
\resizebox{\columnwidth}{!}{%
\begin{tabular}{@{}c|l|cccccc@{}}
\toprule
                        & \multicolumn{1}{c|}{}                                          & \multicolumn{3}{c|}{CVC15}                                                                                                   & \multicolumn{3}{c}{CVC5}                                                                                                     \\ \cmidrule(l){3-8} 
\multirow{-2}{*}{Scale} & \multicolumn{1}{c|}{\multirow{-2}{*}{Methods}}                 & PSNR$\uparrow $                          & MSE$\downarrow $                        & \multicolumn{1}{c|}{SSIM$\uparrow $}    & PSNR$\uparrow $                          & MSE$\downarrow $                        & SSIM$\uparrow $                         \\ \midrule
                        & Shuffle (base)\textcolor[RGB]{217,205,144}{\textit{[NIPS'22]}}~\cite{sun2022shufflemixer} & 42.0601                                  & 4.7414                                  & 0.9634                                  & 41.7926                                  & 4.5484                                  & 0.9696                                  \\
                        & Shuffle (tiny)\textcolor[RGB]{217,205,144}{\textit{[NIPS'22]}}~\cite{sun2022shufflemixer} & 42.1647                                  & 4.6198                                  & \textbf{0.9640}                         & 41.9064                                  & 4.4140                                  & \textbf{0.9701}                         \\
                        & HAT \textcolor[RGB]{217,205,144}{\textit{[CVPR 2023]}}~\cite{chen2023activating}         & {\ul 42.7350}                            & {\ul 3.4555}                            & 0.9633                                  & {\ul 42.5018}                            & {\ul 3.2517}                            & 0.9697                                  \\
                        & SAFMN\textcolor[RGB]{217,205,144}{\textit{[ICCV 2023]}}~\cite{sun2023spatially}        & 35.7437                                  & -                                       & 0.9523                                  & 34.9967                                  & -                                       & 0.9583                                  \\
\multirow{-5}{*}{$\times 2$}    & \cellcolor[HTML]{EFEFEF}Ours                                   & \cellcolor[HTML]{EFEFEF}\textbf{42.7978} & \cellcolor[HTML]{EFEFEF}\textbf{3.4066} & \cellcolor[HTML]{EFEFEF}{\ul 0.9638}    & \cellcolor[HTML]{EFEFEF}\textbf{42.5404} & \cellcolor[HTML]{EFEFEF}\textbf{3.2298} & \cellcolor[HTML]{EFEFEF}{\ul 0.9700}    \\ \midrule
                        & Shuffle (base)\textcolor[RGB]{217,205,144}{\textit{[NIPS'22]}}~\cite{sun2022shufflemixer} & 38.7197                                  & 8.1912                                  & 0.9357                                  & {\ul 38.5379}                            & {\ul 8.2922}                            & {\ul 0.9439}                            \\
                        & Shuffle (tiny)\textcolor[RGB]{217,205,144}{\textit{[NIPS'22]}}~\cite{sun2022shufflemixer} & 38.7721                                  & 8.0796                                  & {\ul 0.9358}                            & 38.5120                                  & 8.3231                                  & 0.9438                                  \\
                        & HAT \textcolor[RGB]{217,205,144}{\textit{[CVPR 2023]}}~\cite{chen2023activating}         & {\ul 38.7845}                            & {\ul 8.0508}                            & {\ul 0.9358}                            & 38.5250                                  & 8.3395                                  & 0.9437                                  \\
                        & SAFMN\textcolor[RGB]{217,205,144}{\textit{[ICCV 2023]}}~\cite{sun2023spatially}        & 29.9648                                  & -                                       & 0.9113                                  & 28.2868                                  & -                                       & 0.9196                                  \\
\multirow{-5}{*}{$\times 4$}    & \cellcolor[HTML]{EFEFEF}Ours                                   & \cellcolor[HTML]{EFEFEF}\textbf{38.9397} & \cellcolor[HTML]{EFEFEF}\textbf{7.7844} & \cellcolor[HTML]{EFEFEF}\textbf{0.9378} & \cellcolor[HTML]{EFEFEF}\textbf{38.7182} & \cellcolor[HTML]{EFEFEF}\textbf{7.9386} & \cellcolor[HTML]{EFEFEF}\textbf{0.9462} \\ \bottomrule
\end{tabular}%
}
\end{table}

\subsection{Ablation Study}

\begin{table}[t]
\caption{\prr{An ablation study was conducted to assess the impact of domain adaptation from visible images, with an upsampling factor of 4, on the M3FD15 and M3FD20 test datasets.} Quantitative results of PSNR without visible light domain adaptation and adaptation results with pretraining are shown separately \textbf{(w/o \& w).}}
\label{tab:domain adaptation}
\resizebox{\columnwidth}{!}{%
\begin{tabular}{@{}c|cccc@{}}
\toprule
\textbf{Dataset} & PSRGAN~\cite{huang2021infrared} $\times 2$        & SRGAN~\cite{ledig2017photo} $\times 2$       & ESRGAN~\cite{wang2018esrgan} $\times 2$     & Ours $\times 2$       \\ \midrule
M3FD15  & 47.02/46.87 & 45.21/46.12 & 45.77/45.86 & 47.02/48.84 \textcolor[RGB]{217,205,144}{\textbf{[1.82dB $\uparrow$]}} \\
M3FD20  & 46.55/45.02 & 45.56/46.50 & 46.17/46.36 & 47.21/48.79 \textcolor[RGB]{217,205,144}{\textbf{[1.58dB $\uparrow$]}} \\ \bottomrule
\end{tabular}%
}
\end{table}

\begin{table}[htbp]
\centering
\caption{Structural ablation study. From left to right, modules are added, and upsampling factors for $\times 2$ \& $\times 4$ are shown ($\times 2$ $\mid $$\times 4$).}
\label{tab:structural ablation}
\resizebox{\columnwidth}{!}{%
\begin{tabular}{@{}c|c|cccc@{}}
\toprule
                                                       \textbf{Dataset}                 & \textbf{Metrics $\uparrow$} & \textbf{Pre train} $\to$ & + $\mathbb{D}_{s_{\text {trans }}}$ $\to$ & + $\mathcal{L}_n^{\mathbb{G}}$ $\to$    & + $\mathcal{L}_{\text {trans }}$             \\ \midrule
\multicolumn{1}{c|}{}                         & \multicolumn{1}{c|}{PSNR} & 47.02/40.11   & 48.10/41.63   & 48.20/41.63   & 48.84/42.11   \\
\multicolumn{1}{c|}{\multirow{-2}{*}{M3FD15}} & \multicolumn{1}{c|}{SSIM} & 0.9900/0.8682 & 0.9903/0.9723 & 0.9913/0.9723 & 0.9915/0.9724 \\ \midrule
\multicolumn{1}{c|}{}                         & \multicolumn{1}{c|}{PSNR} & 47.21/40.83   & 47.42/41.77   & 48.06/41.77   & 48.79/42.31   \\
\multicolumn{1}{c|}{\multirow{-2}{*}{M3FD20}} & \multicolumn{1}{c|}{SSIM} & 0.9909/0.9744 & 0.9912/0.9754 & 0.9920/0.9755 & 0.9919/0.9755 \\ \bottomrule

\end{tabular}%
}
\end{table}

\begin{table}[htbp]
\centering
\caption{Prior ablation studies. From left to right: shallow, middle, and deep priories were selected separately (PSNR/SSIM).}
\label{tab:prior table}
\resizebox{\columnwidth}{!}{%
\begin{tabular}{@{}c|cccccc@{}}
\toprule
\multirow{2}{*}{Dataset} & \multicolumn{2}{c|}{Shallow Priori}               & \multicolumn{2}{c|}{\textbf{Middle Priori}}     & \multicolumn{2}{c}{Deep Priori}        \\ \cmidrule(l){2-7} 
                         & \multicolumn{1}{c|}{$\times 2$} & \multicolumn{1}{c|}{$\times 4$} & \multicolumn{1}{c|}{$\times 2$} & \multicolumn{1}{c|}{$\times 4$} & \multicolumn{1}{c|}{$\times 2$} & $\times 4$           \\ \midrule
M3FD15                   & 47.21/0.9863            & 41.23/0.9622            & 48.01/0.9889            & 41.66/0.9712            & 47.33/0.9871            & 41.36/0.9672 \\ \cmidrule(r){1-1}
M3FD20                   & 46.92/0.9811            & 41.11/0.9598            & 47.32/0.9901            & 41.53/0.9688            & 47.32/0.9901            & 41.13/0.9602 \\ \bottomrule
\end{tabular}%
}
\end{table}

\begin{table}[htbp]
\centering
\caption{Hyperparameter ablation study  $\times 2$, where $\alpha$ is the weight of $\mathcal{L}_n^{\mathbb{G}}$ and $\beta$ is the weight of $\mathcal{L}_{\text {trans }}$.}
\label{tab:Hyperparameter}
\resizebox{0.65\columnwidth}{!}{%
\begin{tabular}{@{}c|ccc@{}}
\toprule
$\alpha$ /$\beta $    & M3FD5        & M3FD15       & M3FD20       \\ \midrule
0.1/0.0 & 48.32/0.9922 & 48.75/0.9914 & 48.71/0.9919 \\
0.0/1.0 & 48.84/0.9915 & 48.84/0.9900 & 48.79/0.9918 \\
0.5/1.0 & 44.19/0.9920 & 44.65/0.9914 & 45.06/0.9919 \\
0.5/0.1 & 47.93/0.9923 & 48.17/0.9913 & 48.15/0.9920 \\
1.0/0.0 & 39.99/0.9504 & 41.63/0.9723 & 41.77/0.9755 \\
\rowcolor[HTML]{EFEFEF} 
0.1/1.0 & 48.38/0.9922 & 48.84/0.9915 & 48.79/0.9919 \\ \bottomrule
\end{tabular}
}
\end{table}
\vspace{-0.2cm}

\begin{figure}[htbp]
    \centering
    \resizebox{0.9\textwidth}{!}{%
    \includegraphics{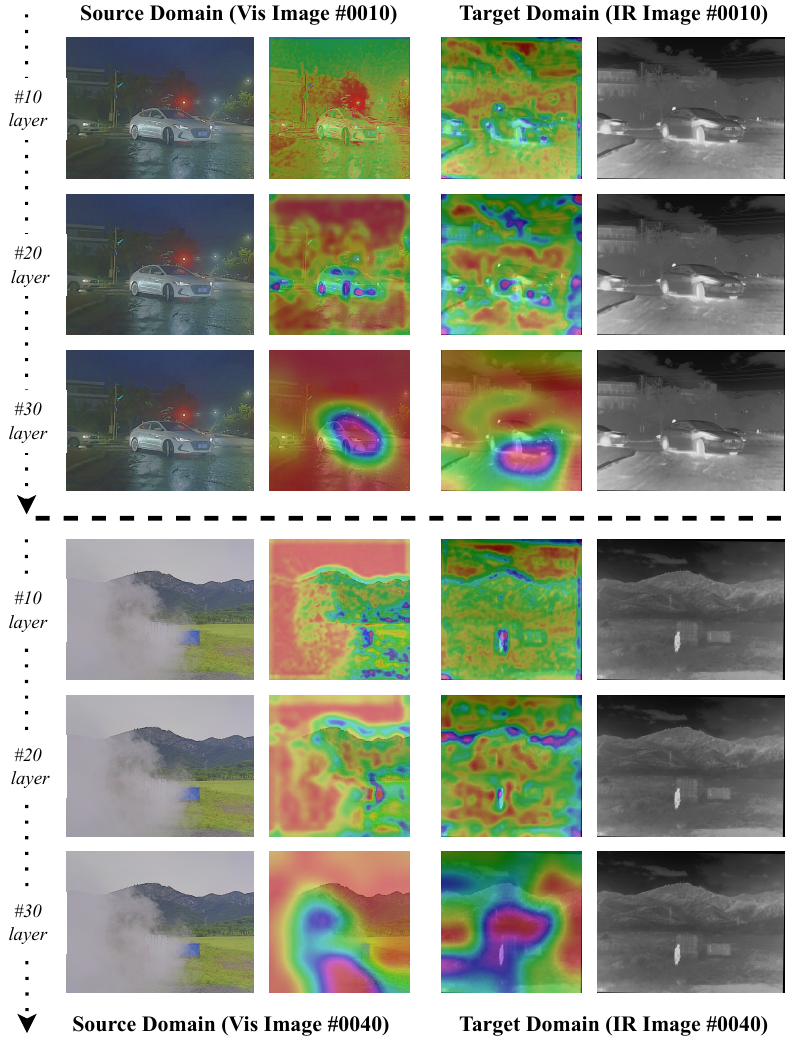}
    }
    \caption{\pr{In visualizing infrared IR images, feature representations are observed in network layers with different depths. Deep priors, focusing on high-level semantics, are ideal for object detection. Meanwhile, mid-level priors, which capture background textures, enhance detail recovery in IR imagery.} In each group, column 1: visible image; column 2: the feature map of the selected layer (Vis Image); column 3: the feature map of the selected layer (IR Image); column 4: IR Image.}
    \label{visualization}
\end{figure}

\textbf{Effectiveness of the domain adaptation.} The domain adaptation strategy is beneficial to the PSNR metrics. However, the performance of PSRGAN~\cite{huang2021infrared}, a model sensitive to negative features (\eg, noise), declines if the feature domains are directly transferred. DASRGAN's \prrr{Texture-noise Oriented} domain adaptation strategy achieves better performance than its competitors (see Tab.\ref{tab:domain adaptation}). \textbf{Effectiveness of Texture-noise Oriented.}
To rigorously assess the efficacy of the proposed DASRGAN model, we conducted a series of preliminary experiments.
The results show consistent gains with the stepwise addition of $\mathbb{D}_{s_{\text {trans }}} \rightarrow +\mathcal{L}_n^{\mathbb{G}} \rightarrow +\mathcal{L}_{\text {trans }}$. This confirms the utility of both a priori and noise-based domain adaptation in SR tasks for IR images. For detailed metrics, see Tab.\ref{tab:structural ablation}. \textbf{Effectiveness of Prior.}
To evaluate the impact of texture priors across feature layers, we compare results in Tab.\ref{tab:prior table}. The data show that introducing texture prior to the middle layer is most effective in enhancing IR image details. Deep priors focus on high-level semantics, making them ideal for object detection, while mid priors better capture background texture, optimizing detail recovery in IR images (see Fig.\ref{visualization}). \textbf{Hyper-parameter Ablation} In the hyperparameter ablation study, the optimal setting is found to be \( \alpha = 0.1 \) and \( \beta = 1.0 \), delivering the highest performance across all metrics. Conversely, a configuration with \( \alpha = 1.0 \) and \( \beta = 0.0 \) results in the poorest performance. The model demonstrates greater sensitivity to \( \beta \) than \( \alpha \) (see Tab.\ref{tab:Hyperparameter}). \hyspr{We tested three different learning rates: $1e-3,1e-4$, and $1e-5$, across both Stage 1 and Stage 2 of the DASRGAN framework. The pre-trained model from Stage 1 was used as the initialization for Stage 2, maintaining the same learning rate across stages for each configuration(see Tab.\ref{Stage Ablation}). In stage 1, the learning rate of $1 e-4$ provided a balance between performance and training efficiency, yielding a PSNR of 41.76 dB and an SSIM of 0.9709 in the M3FD15 data set. Similarly, it achieved a PSNR of 42.57 dB and an SSIM of 0.9739 on the M3FD20 dataset. This learning rate offered superior results compared to $1 e-3$ and comparable performance to $1 e-5$, while requiring less computational time. In stage 2, we observed that the learning rate of $1 e-5$ provided the best overall performance, with the highest PSNR and SSIM across both datasets. Specifically, it achieved a PSNR of 43.08 dB on M3FD15 and 43.57 dB on M3FD20, with SSIM values of 0.9722 and 0.9750, respectively. Based on these results, we selected the learning rate of $1e-5$ for the final model, as it provided the best balance between training stability and overall performance in both stages.}

\begin{table}[htbp]
\caption{Comparison of learning rates in Stage 1 and Stage 2 at $4\times$  upsampling. The pre-trained model for Stage 2 is derived from Stage 1 using the same learning rate (PSNR/SSIM).}
\centering
\label{table1}
\begin{tabular}{@{}cccc@{}}
\toprule
Scale $4\times$                                      & Learning Rates            & M3FD15       & M3FD20       \\ \midrule
\multicolumn{1}{c|}{\multirow{3}{*}{Stage 1}} & \multicolumn{1}{c|}{1e-3} & 41.09/0.9610 & 42.32/0.9643 \\
\multicolumn{1}{c|}{}                         & \multicolumn{1}{c|}{1e-4} & 41.76/0.9709 & 42.57/0.9739 \\
\multicolumn{1}{c|}{}                         & \multicolumn{1}{c|}{1e-5} & 42.03/0.9713 & 42.45/0.9730 \\ \midrule
\multicolumn{1}{c|}{\multirow{3}{*}{Stage 2}} & \multicolumn{1}{c|}{1e-3} & 42.36/0.9756 & 43.05/0.9707 \\
\multicolumn{1}{c|}{}                         & \multicolumn{1}{c|}{1e-4} & 42.54/0.9731 & 43.36/0.9764 \\
\multicolumn{1}{c|}{}                         & \multicolumn{1}{c|}{1e-5} & 43.08/0.9722 & 43.57/0.9750 \\ \bottomrule
\end{tabular}
\label{Stage Ablation}
\end{table}

\textbf{Domain adaptation ablation experiment.}
Our study employed a domain adaptation ablation experiment to assess the effectiveness of direct domain adaptation against our innovative domain adaptation model. This experiment was crucial to highlight the contrast between our approach and the conventional methods used by our competitors. Typically, competing models are trained on visible light data, followed by a straightforward adaptation to other domains.
The experiment featuring DASRGAN focused on addressing key challenges in IR image super-resolution (see Tab.\ref{tab2}), with an emphasis on domain alignment. Traditional super-resolution methods tend to amplify texture details in visible light images but often introduce unwanted noise and blurring. DASRGAN, leveraging its novel TOA and NOA, aimed to achieve an optimal balance between enhancing texture and minimizing noise, a critical factor for superior quality in IR image super-resolution.

\begin{table*}[htbp]
\caption{Domain adaptation ablation quantitative comparison (PSNR$\uparrow$ MSE$\downarrow$ SSIM$\uparrow$). Best performances are marked in \textbf{bold}.}
\label{tab2}
\resizebox{\textwidth}{!}{%
\begin{tabular}{@{}c|c|c|cccccccccccc@{}}
\toprule
\multirow{2}{*}{Scale} & \multirow{2}{*}{Methods} & \multirow{2}{*}{\#Params} & \multicolumn{3}{c|}{M3FD20}                                    & \multicolumn{3}{c|}{M3FD15}                                    & \multicolumn{3}{c|}{M3FD5}                                                       & \multicolumn{3}{c}{CVC15}                            \\ \cmidrule(l){4-15} 
                       &                          &                           & PSNR$\uparrow $              & MSE$\downarrow $             & \multicolumn{1}{c|}{SSIM$\uparrow $ } & PSNR$\uparrow $              & MSE$\downarrow $             & \multicolumn{1}{c|}{SSIM$\uparrow $ } & PSNR$\uparrow $  & MSE$\downarrow $ & \multicolumn{1}{c|}{SSIM$\uparrow $ } & PSNR$\uparrow $            & MSE$\downarrow $             & SSIM$\uparrow $             \\ \midrule
\multirow{5}{*}{$\times 2$}    & Shuffle (tiny)\textcolor[RGB]{217,205,144}{\textit{[NIPS'22]}}\cite{sun2022shufflemixer}          & 108K                      & 46.6025          & 1.2430          & 0.9912                    & 46.0969          & 1.5387          & 0.9903                    & 45.7881                   & 2.3947                   & 0.9893                    & 42.2029          & 4.0526          & 0.9708          \\
                       & Shuffle (base)\textcolor[RGB]{217,205,144}{\textit{[NIPS'22]}}\cite{sun2022shufflemixer}           & 393K                      & 46.5929          & 1.2406          & 0.9913                    & 46.1603          & 1.5251          & 0.9903                    & 45.7532                   & 2.3918                   & 0.9894                    & 42.2286          & 4.0184          & 0.9708          \\
                       &  HAT \textcolor[RGB]{217,205,144}{\textit{[CVPR 2023]}}\cite{chen2023activating}                     & 20,624K                   & {\ul 47.3946}    & {\ul 0.8919}    & {\ul 0.9915}              & {\ul 46.9474}    & {\ul 1.0858}    & {\ul 0.9906}              & {\ul 46.5997}             & {\ul 1.6941}             & {\ul 0.9897}              & \textbf{42.9901} & \textbf{2.8610} & \textbf{0.9712} \\
                       &  SAFMN\textcolor[RGB]{217,205,144}{\textit{[ICCV 2023]}}\cite{sun2023spatially}                   & 5,559K                    & 46.9984          & 0.9798          & {\ul 0.9915}              & 46.4638          & 1.2035          & {\ul 0.9906}              & 46.2635                   & 1.7728                   & 0.9896                    & {\ul 42.8808}    & {\ul 2.9359}    & {\ul 0.9709}    \\
                       & Ours                     & 16,661K                   & \textbf{48.7952} & \textbf{0.6410} & \textbf{0.9919}           & \textbf{48.8402} & \textbf{0.7148} & \textbf{0.9915}           & \textbf{48.3865}          & \textbf{0.9982}          & \textbf{0.9922}           & 42.5404          & 3.2298          & 0.9700          \\ \midrule
\multirow{5}{*}{$\times 4$}    & Shuffle (tiny)\textcolor[RGB]{217,205,144}{\textit{[NIPS'22]}}\cite{sun2022shufflemixer}           & 108K                      & 40.0529          & 5.2913          & 0.9708                    & 39.2513          & 6.3910          & 0.9674                    & 37.8134                   & 14.1767                  & 0.9442                    & 36.6736          & 12.8571         & 0.9240          \\
                       & Shuffle (base)\textcolor[RGB]{217,205,144}{\textit{[NIPS'22]}}\cite{sun2022shufflemixer}          & 393K                      & 40.7455          & 4.4358          & 0.9731                    & 40.2344          & 5.0625          & 0.9700                    & 38.5625                   & 11.5552                  & 0.9503                    & 37.7422          & 10.0539         & 0.9318          \\
                       &  HAT \textcolor[RGB]{217,205,144}{\textit{[CVPR 2023]}}\cite{chen2023activating}                      & 20,624K                   & {\ul 41.1975}    & {\ul 3.8931}    & 0.9730                    & 40.6626          & 4.5943          & 0.9697                    & {\ul 39.1975}             & {\ul 10.6138}            & 0.9505                    & {\ul 38.3079}    & {\ul 8.8542}    & {\ul 0.9334}    \\
                       &  SAFMN\textcolor[RGB]{217,205,144}{\textit{[ICCV 2023]}}\cite{sun2023spatially}                    & 5,559K                    & 40.9292          & 4.2033          & {\ul 0.9734}              & {\ul 40.5122}    & {\ul 4.7797}    & {\ul 0.9704}              & 38.8705                   & 11.0464                  & {\ul 0.9510}              & 38.2170          & 9.0804          & 0.9341          \\
                       & Ours                     & 16,661K                   & \textbf{42.3179} & \textbf{3.0009} & \textbf{0.9755}           & \textbf{42.1196} & \textbf{3.3763} & \textbf{0.9724}           & \textbf{40.5088}          & \textbf{8.7855}          & \textbf{0.9558}           & \textbf{38.9397} & \textbf{7.7844} & \textbf{0.9378} \\ \bottomrule
\end{tabular}%
}
\end{table*}

\textbf{Noise-Oriented ablation experiment.} The purpose of the domain adaptation ablation experiment incorporating NOA into DASRGAN was to assess the effectiveness of NOA in enhancing the model's denoising capabilities, particularly against images affected by different levels of Gaussian blur. This was achieved by applying Gaussian blur kernels with standard deviations ($\sigma$) of 1, 3, and 5, to test the model's adaptability in reducing noise under varying blur intensities (see Tab.\ref{tabl2}).

\begin{table}[htbp]
\centering
\caption{Gaussian ablation study. From left to right, the Gaussian blur kernel $\sigma$ is 1, 3, and 5 respectively (PSNR/SSIM).}
\label{tabl2}
\resizebox{\textwidth}{!}{%
\begin{tabular}{@{}cc|ccc@{}}
\toprule
                                                       & \textbf{Method}              & \multicolumn{1}{c|}{\includegraphics[width=0.1\columnwidth]{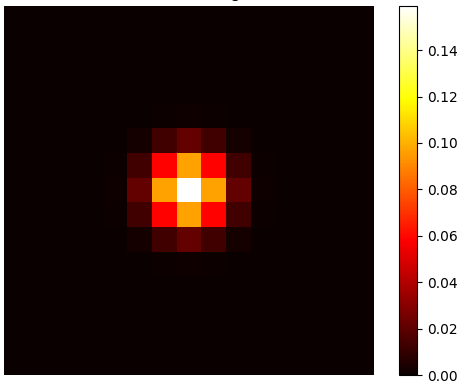}}                & \multicolumn{1}{c|}{{\includegraphics[width=0.1\columnwidth]{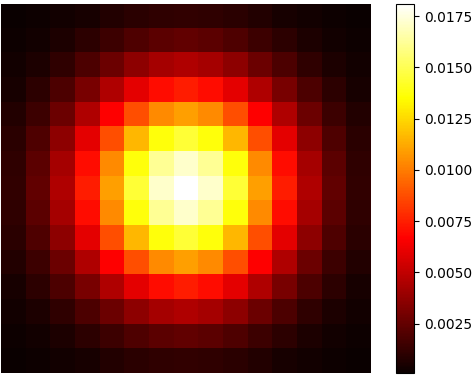}}}                & {\includegraphics[width=0.1\columnwidth]{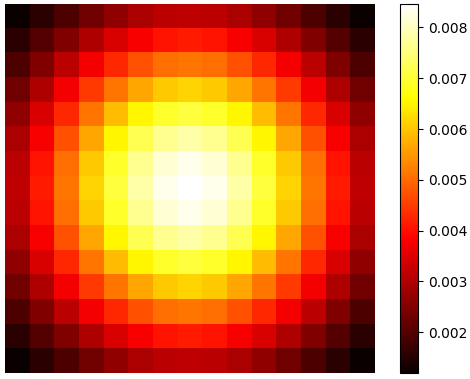}}                                     \\ \midrule
\multicolumn{1}{c|}{}                                  & HAT                          & 36.5357/0.9624                         & 30.6954/0.9291                         & 29.2792/0.9214                         \\
\multicolumn{1}{c|}{}                                  & SAFMN                        & 36.5151/0.9623                         & 30.5270/0.9276                         & 28.9087/0.9195                         \\
\multicolumn{1}{c|}{\multirow{-3}{*}{\textbf{M3FD15}}} & \cellcolor[HTML]{EFEFEF}Ours & \cellcolor[HTML]{EFEFEF}37.5374/0.9626 & \cellcolor[HTML]{EFEFEF}36.6965/0.9291 & \cellcolor[HTML]{EFEFEF}29.2791/0.9214 \\ \midrule
\multicolumn{1}{c|}{}                                  & HAT                          & 35.6671/0.9333                         & 29.4772/0.8565                         & 28.0796/0.8420                         \\
\multicolumn{1}{c|}{}                                  & SAFMN                        & 35.6860/0.9335                         & 29.3604/0.8553                         & 27.8046/0.8398                         \\
\multicolumn{1}{c|}{\multirow{-3}{*}{\textbf{M3FD5}}}  & \cellcolor[HTML]{EFEFEF}Ours & \cellcolor[HTML]{EFEFEF}35.6663/0.9334 & \cellcolor[HTML]{EFEFEF}29.4778/0.8566 & \cellcolor[HTML]{EFEFEF}28.0791/0.8420 \\ \bottomrule
\end{tabular}%
}
\end{table}

\begin{figure}[htbp]
    \centering
    \resizebox{\textwidth}{!}{%
    \includegraphics{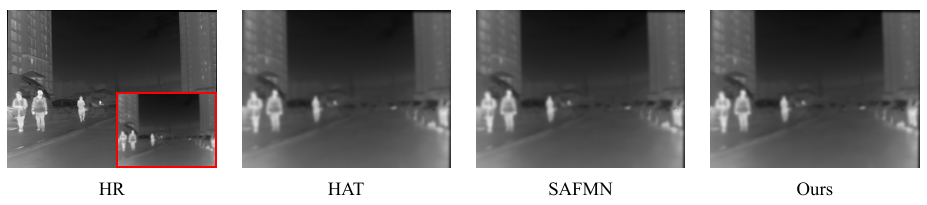}
    }
    \caption{Visual comparison of $\times 4$ Gaussian-added image super-resolution methods (M3FD5 Dataset).}
    \label{fig2}
\end{figure}

In our experiment, we assessed the performance of DASRGAN, specifically its NOA feature, in comparison with leading methods like HAT and SAFMN. We used two datasets for this evaluation: M3FD15 and M3FD5. The performance metrics focused on PSNR and SSIM.

The results were particularly noteworthy in scenarios involving higher blur levels ($\sigma$ = 3 and 5), especially with the M3FD15 dataset. In these instances, DASRGAN with NOA exhibited a remarkable performance, significantly outperforming the other methods. This was reflected in the higher PSNR and SSIM scores achieved by our method, signaling an enhanced quality of image and structural integrity even under conditions of substantial noise (see Fig.\ref{fig2}).

In the case of the M3FD5 dataset, while the improvements by our method were less pronounced, it still maintained competitive denoising capabilities, especially at higher blur levels. This consistent performance across various degrees of Gaussian blur underscores the effectiveness of the NOA component in DASRGAN. The integration of NOA has evidently fortified the model's robustness, enabling it to adeptly manage different noise patterns and blur intensities. This enhanced adaptability of DASRGAN with NOA is a significant advancement, demonstrating its potential for widespread application in scenarios where handling varying levels of noise and blur is critical.

\begin{table*}[]
\caption{Priors operator ablation quantitative comparison (PSNR$\uparrow$ MSE$\downarrow$ SSIM$\uparrow$). Best performances are marked in \textbf{bold}.}
\label{tab3}
\resizebox{\textwidth}{!}{%
\begin{tabular}{@{}c|cccccccccccccccc@{}}
\toprule
                        & \multicolumn{1}{c|}{}                                  & \multicolumn{3}{c|}{M3FD5}                                                                                                   & \multicolumn{3}{c|}{M3FD15}                                                                                                  & \multicolumn{3}{c|}{M3FD20}                                                                                                  & \multicolumn{3}{c|}{CVC5}                                                                                           & \multicolumn{3}{c}{CVC15}                                                                                           \\ \cmidrule(l){3-17} 
\multirow{-2}{*}{scale} & \multicolumn{1}{c|}{\multirow{-2}{*}{Priors operator}} & \multicolumn{1}{c|}{PSNR}                & \multicolumn{1}{c|}{MSE}                & \multicolumn{1}{c|}{SSIM}               & \multicolumn{1}{c|}{PSNR}                & \multicolumn{1}{c|}{MSE}                & \multicolumn{1}{c|}{SSIM}               & \multicolumn{1}{c|}{PSNR}                & \multicolumn{1}{c|}{MSE}                & \multicolumn{1}{c|}{SSIM}               & \multicolumn{1}{c|}{PSNR}                & \multicolumn{1}{c|}{MSE}                & \multicolumn{1}{c|}{SSIM}      & \multicolumn{1}{c|}{PSNR}                & \multicolumn{1}{c|}{MSE}                & SSIM                           \\ \midrule
                        & \multicolumn{1}{c|}{Laplacian}                         & 47.4833                                  & 1.2932                                  & 0.9910                                  & 48.0262                                  & 0.8688                                  & 0.9908                                  & 48.2071                                  & 0.7356                                  & 0.9916                                  & 42.3852                                  & 3.3352                                  & 0.9702                         & 42.6384                                  & 3.5039                                  & 0.9640                         \\
                        & \multicolumn{1}{c|}{Scharr}                            & 47.0151                                  & 1.4646                                  & 0.9903                                  & 47.4984                                  & 0.9653                                  & 0.9906                                  & 47.7160                                  & 0.8236                                  & 0.9915                                  & 42.4766                                  & -                                       & \textbf{0.9706}                & 42.6204                                  & -                                       & \textbf{0.9644}                \\
\multirow{-3}{*}{$\times 2$}    & \multicolumn{1}{c|}{\cellcolor[HTML]{C0C0C0}Ours}      & \cellcolor[HTML]{C0C0C0}\textbf{48.3865} & \cellcolor[HTML]{C0C0C0}\textbf{0.9982} & \cellcolor[HTML]{C0C0C0}\textbf{0.9922} & \cellcolor[HTML]{C0C0C0}\textbf{48.8402} & \cellcolor[HTML]{C0C0C0}\textbf{0.7148} & \cellcolor[HTML]{C0C0C0}\textbf{0.9915} & \cellcolor[HTML]{C0C0C0}\textbf{48.7952} & \cellcolor[HTML]{C0C0C0}\textbf{0.6410} & \cellcolor[HTML]{C0C0C0}\textbf{0.9919} & \cellcolor[HTML]{C0C0C0}\textbf{42.5404} & \cellcolor[HTML]{C0C0C0}\textbf{3.2298} & \cellcolor[HTML]{C0C0C0}0.9700 & \cellcolor[HTML]{C0C0C0}\textbf{42.7978} & \cellcolor[HTML]{C0C0C0}\textbf{3.4066} & \cellcolor[HTML]{C0C0C0}0.9638 \\ \midrule
                        & \multicolumn{1}{c|}{Laplacian}                         & 39.2961                                  & 10.4237                                 & 0.9525                                  & 40.6064                                  & 4.6815                                  & 0.9712                                  & 41.2124                                  & 3.8618                                  & 0.9743                                  & \textbf{39.4126}                         & -                                       & \textbf{0.9492}                & \textbf{39.6205}                         & \textbf{6.7619}                         & \textbf{0.9408}                \\
                        & \multicolumn{1}{c|}{Scharr}                            & 38.2379                                  & 14.3234                                 & 0.9474                                  & 39.7535                                  & 5.8710                                  & 0.9656                                  & 40.0628                                  & 5.1932                                  & 0.9686                                  & 37.2833                                  & 10.9895                                 & 0.9405                         & 37.4671                                  & 10.6918                                 & 0.9319                         \\
\multirow{-3}{*}{$\times 4$}    & \cellcolor[HTML]{C0C0C0}Ours                           & \cellcolor[HTML]{C0C0C0}\textbf{40.5088} & \cellcolor[HTML]{C0C0C0}\textbf{8.7855} & \cellcolor[HTML]{C0C0C0}\textbf{0.9558} & \cellcolor[HTML]{C0C0C0}\textbf{42.1196} & \cellcolor[HTML]{C0C0C0}\textbf{3.3763} & \cellcolor[HTML]{C0C0C0}\textbf{0.9724} & \cellcolor[HTML]{C0C0C0}\textbf{42.3179} & \cellcolor[HTML]{C0C0C0}\textbf{3.0009} & \cellcolor[HTML]{C0C0C0}\textbf{0.9755} & \cellcolor[HTML]{C0C0C0}38.7182          & \cellcolor[HTML]{C0C0C0}\textbf{7.9386} & \cellcolor[HTML]{C0C0C0}0.9462 & \cellcolor[HTML]{C0C0C0}38.9397          & \cellcolor[HTML]{C0C0C0}7.7844          & \cellcolor[HTML]{C0C0C0}0.9378 \\ \bottomrule
\end{tabular}%
}
\end{table*}

\begin{figure*}[t]
    \centering
    \resizebox{\textwidth}{!}{%
    \includegraphics{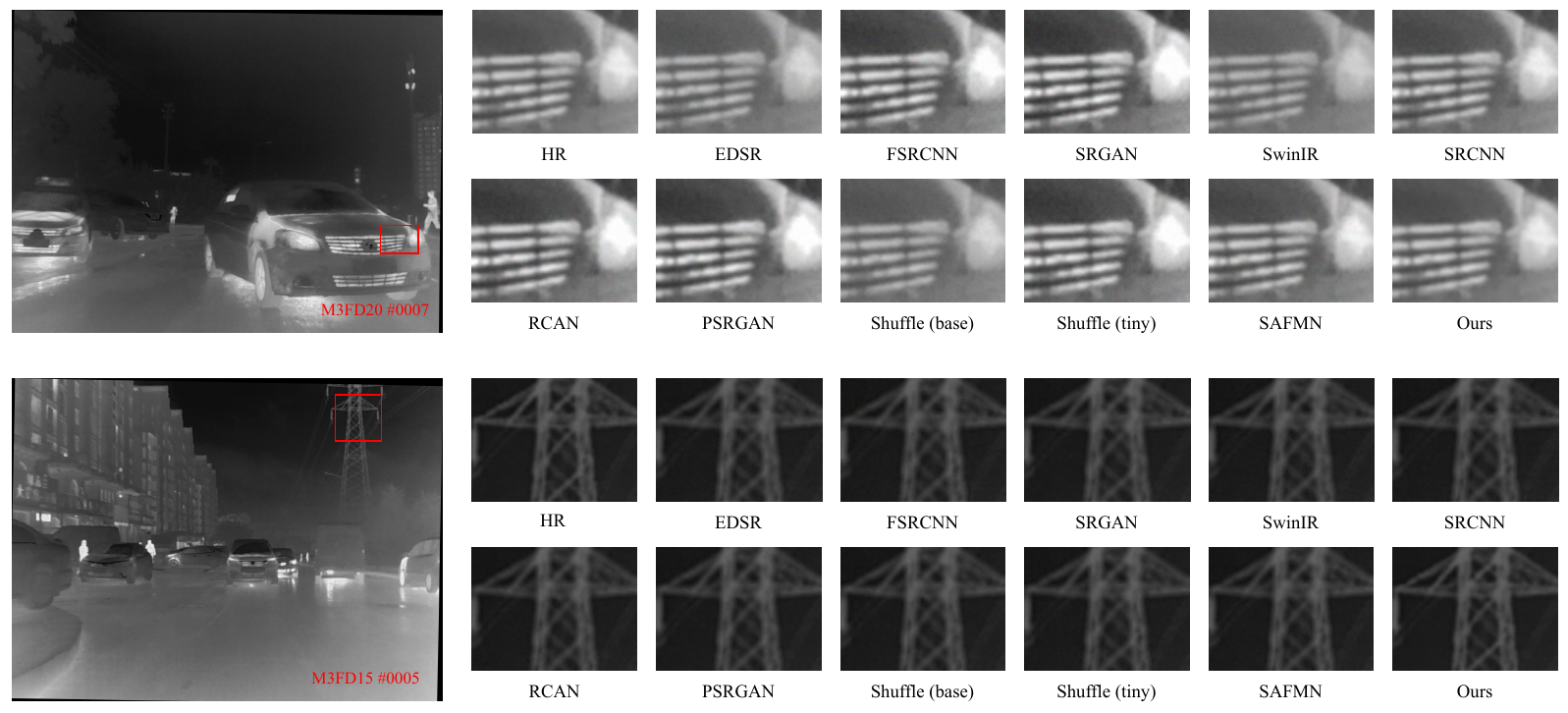}
    }
    \caption{Visual comparison of $\times 2$ image super-resolution methods (M3FD Datasets). Zoom-in for better details.}
    \label{fig:m3fd x2}
\end{figure*}

\begin{figure*}[t]
    \centering
    \resizebox{\textwidth}{!}{%
    \includegraphics{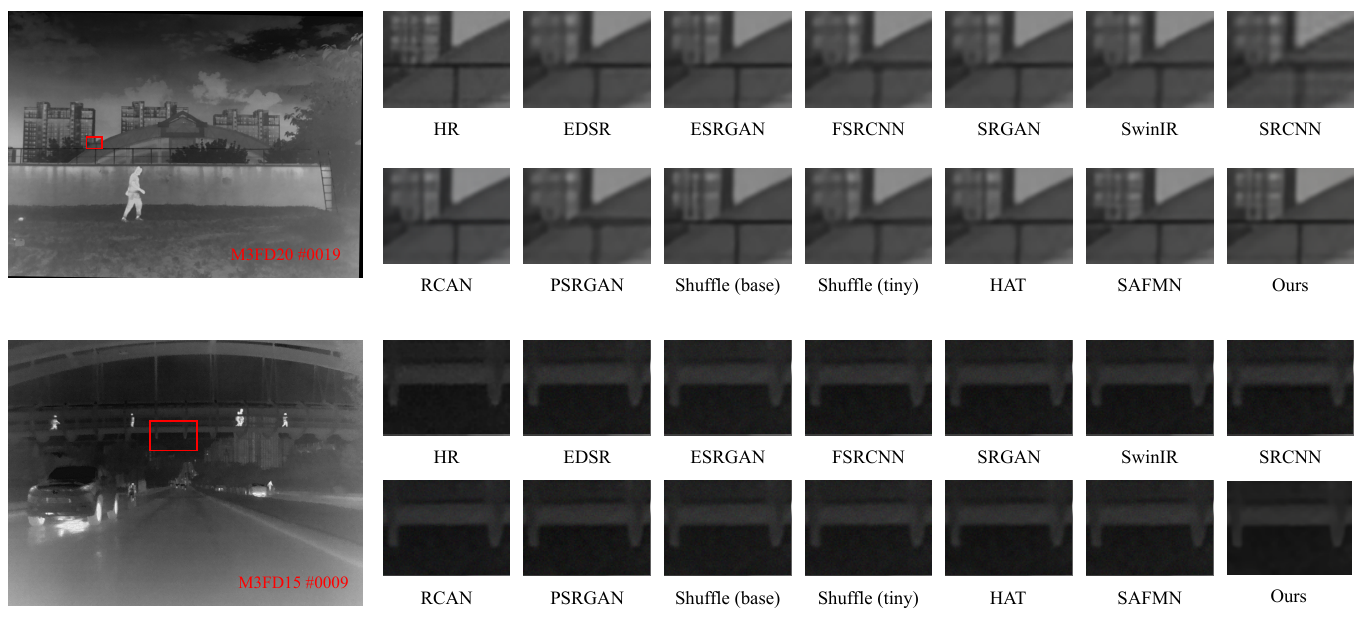}
    }
    \caption{Visual comparison of $\times 4$ image super-resolution methods (M3FD Datasets). Zoom-in for better details.}
    \label{fig:m3fd x4}
\end{figure*}

\textbf{Priors operator ablation experiment} The ablation experiment with a focus on TOA in the DASRGAN model was designed to assess the impact of different prior operators on the reconstruction performance across various datasets (see Tab.\ref{tab3}). The experiment compared the performance of the model using different prior operators, specifically Laplacian, Scharr, and the operator used in our method (Ours), across datasets M3FD5, M3FD15, M3FD20, CVC5, and CVC15. 

At the ×2 scale, our method consistently surpassed the performance of both the Laplacian and Scharr operators across all datasets. This superiority was particularly evident in the M3FD20 dataset, where our method not only achieved the highest PSNR but also the lowest MSE, affirming its exceptional capability in reconstructing images with enhanced quality. When we moved to the more challenging $\times 4$ scale, our method maintained its superior performance. This was especially noticeable in the M3FD15 and M3FD20 datasets, where it recorded the highest PSNR and the lowest MSE. These results underscore the effectiveness of our method in handling complex image reconstruction tasks at higher scales. In the context of the CVC datasets, our method showcased competitive performance, despite not always achieving the highest SSIM values. Notably, it consistently maintained low MSE scores, a clear indicator of minimal error in the reconstructed images. This aspect is particularly significant as it demonstrates the method's capability to produce high-quality reconstructions with fewer distortions, a critical factor in practical applications.

\begin{figure*}[t]
    \centering
    \resizebox{\textwidth}{!}{%
    \includegraphics{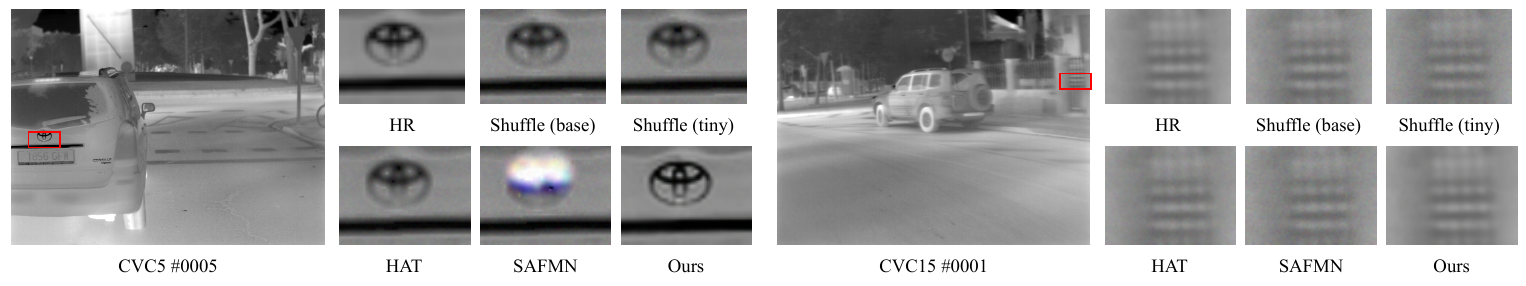}
    }
    \caption{Visual comparison of $\times 2$ image super-resolution methods (CVC Datasets). Zoom-in for better details.}
    \label{fig:cvc x2}
\end{figure*}
\vspace{-0.1cm}

\begin{figure*}[htbp]
    \centering
    \resizebox{\textwidth}{!}{%
    \includegraphics{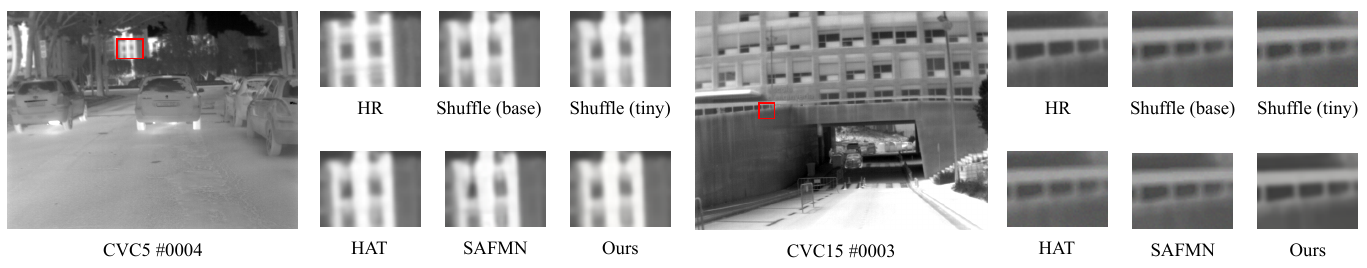}
    }
    \caption{Visual comparison of $\times 4$ image super-resolution methods (CVC Datasets). Zoom-in for better details.}
    \label{fig: cvc x4}
\end{figure*}
\vspace{-0.1cm}

\subsection{Visual Comparison}
In Fig.\ref{fig:m3fd x2} \& Fig.\ref{fig:m3fd x4}, we present a visual comparison featuring images "0007" "0005" "0009" "0019" from the M3FD15 \& M3FD20 dataset. DASRGAN effectively recovers crisp lattice structures, whereas competing methods exhibit pronounced blurring. Similar observations can be made for the "0019" image in the M3FD20 dataset, where DASRGAN excels in restoring clear textures for the characters. \hyspr{In qualitative comparisons, DASRGAN consistently surpasses baseline methods in reducing blur and minimizing artifacts. Notably, approaches such as PSRGAN and RCAN commonly introduce ringing artifacts or fail to recover texture details fully, leaving residual blur in critical image areas (\eg, CVC 0005). These methods face challenges in preserving texture, particularly under varying data distributions, whereas our model maintains both texture fidelity and artifact suppression. This improvement is largely due to the enhanced texture adaptation enabled by our model's architecture, which provides more flexibility in handling noise and texture variations. In addition, our method excels in texture and noise adaptation. Baseline methods, such as PSRGAN and HAT, often introduce artifacts or struggle to fully suppress noise when dealing with complex textures. In contrast, our model effectively adapts to both noise and texture variations across different image regions by strategically incorporating texture priors, which mitigates artifacts and enhances noise suppression without sacrificing detail recovery.}


\begin{figure*}[htbp]
    \centering
    \resizebox{\textwidth}{!}{%
    \includegraphics{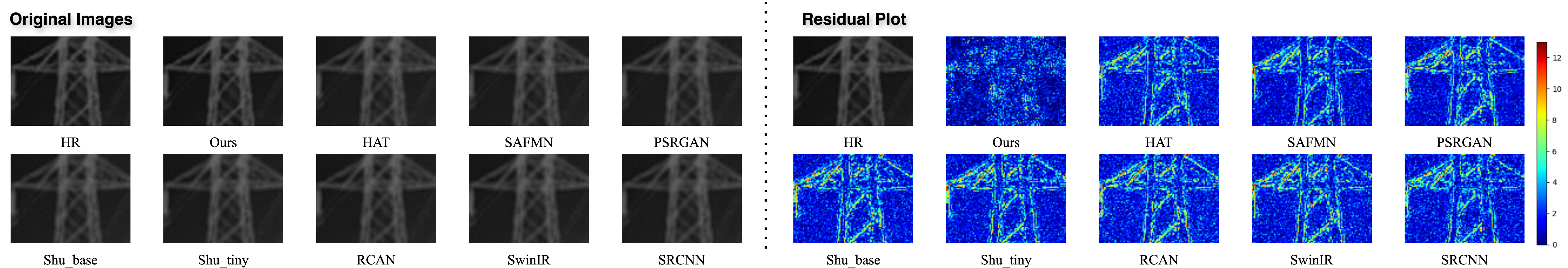}
    }
    \caption{The residual plots of different SR methods compared to the HR image for $\times 2$ upsampling.}
    \label{fig: x2}
\end{figure*}
\vspace{-0.1cm}

\begin{figure*}[htbp]
    \centering
    \resizebox{\textwidth}{!}{%
    \includegraphics{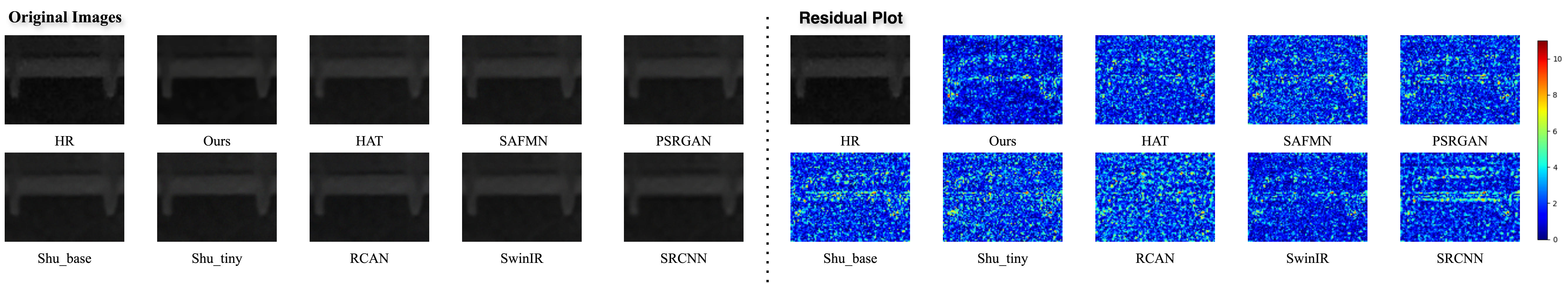}
    }
    \caption{The residual plots of different SR methods compared to the HR image for $\times 4$ upsampling.}
    \label{fig: x4}
\end{figure*}
\vspace{-0.1cm}

\hyspr{Following the visual comparison, we selected two representative SR results to perform residual visualization for better clarity in performance differences. Fig.\ref{fig: x2} and Fig.\ref{fig: x4} illustrate the performance differences between the proposed DASRGAN model and baseline methods in terms of texture preservation and noise suppression. In both the $\times 2$  and $\times 4$ upsampling scenarios, the residual plots depict the magnitude of error between the HR and SR images, where lower residuals indicate better performance. As shown by the color bars, DASRGAN consistently produces lower residuals, especially in critical areas such as edges and textures, where other methods tend to falter. In Fig.\ref{fig: x2}, DASRGAN demonstrates superior texture recovery with notably reduced residuals compared to methods like PSRGAN and HAT, which exhibit higher errors around complex structural details. Similarly, in Fig.\ref{fig: x4}, DASRGAN continues to outperform the baseline methods, maintaining minimal residuals even under the more demanding $\times 4$ upsampling condition, where other models introduce visible artifacts. These findings underscore the effectiveness of the proposed texture-noise adaptation framework in accurately preserving fine details and mitigating noise across different super-resolution scales.}

\section{Conclusion\label{sec4}}
\hyspr{In this study, we introduce Texture and Noise Dual Adaptation SRGAN,
a novel framework specifically designed for IR image super-resolution through visible image feature domain adaptation. The framework is built on two key components: 1) Texture-Oriented Adaptation, which enhances texture fidelity using a specialized prior adversarial loss $\mathcal{L}_{\text {trans }}$ and a tailored discriminator $\mathbb{D}_{s_{\text {trans }}}$ to transfer textural features from visible images, and 2) Noise-Oriented Adaptation, which minimizes noise transfer from the feature domain via an innovative noise adversarial loss $\mathcal{L}_n^{\mathbb{G}}$. Extensive experiments validate the effectiveness of these components, setting new benchmarks in both qualitative and quantitative evaluations. The framework shows strong potential in real-world applications where high-quality IR imagery is essential, such as surveillance, medical imaging, and autonomous driving. Enhanced IR resolution improves object detection in low-light environments, increases diagnostic accuracy by providing clearer thermal properties in medical scans, and aids obstacle detection for safer navigation in self-driving systems, making it valuable for security, healthcare, and automotive industries.}

\newpage

\section*{Declaration of competing interest.} The authors declare that they have no known competing financial interests or personal relationships that could have appeared to
influence the work reported in this paper.

\section*{Data availability.} Data will be made available on request.

\section*{Acknowledgments.} This work was supported by JSPS KAKENHI Grant Number JP23KJ0118. 

\bibliographystyle{elsarticle-harv.bst} 
\bibliography{pr_r}

\begin{thebibliography}{45}
\expandafter\ifx\csname natexlab\endcsname\relax\def\natexlab#1{#1}\fi
\providecommand{\url}[1]{\texttt{#1}}
\providecommand{\href}[2]{#2}
\providecommand{\path}[1]{#1}
\providecommand{\DOIprefix}{doi:}
\providecommand{\ArXivprefix}{arXiv:}
\providecommand{\URLprefix}{URL: }
\providecommand{\Pubmedprefix}{pmid:}
\providecommand{\doi}[1]{\href{http://dx.doi.org/#1}{\path{#1}}}
\providecommand{\Pubmed}[1]{\href{pmid:#1}{\path{#1}}}
\providecommand{\bibinfo}[2]{#2}
\ifx\xfnm\relax \def\xfnm[#1]{\unskip,\space#1}\fi
\bibitem[{Campo et~al.(2012)Campo, Ruiz and Sappa}]{campo2012multimodal}
\bibinfo{author}{Campo, F.B.}, \bibinfo{author}{Ruiz, F.L.}, \bibinfo{author}{Sappa, A.D.}, \bibinfo{year}{2012}.
\newblock \bibinfo{title}{Multimodal stereo vision system: 3d data extraction and algorithm evaluation}.
\newblock \bibinfo{journal}{IEEE Journal of Selected Topics in Signal Processing} \bibinfo{volume}{6}, \bibinfo{pages}{437--446}.
\bibitem[{Chen et~al.(2024)Chen, Chen, Chen, Chen, Sheng, Yu and Zou}]{10379652}
\bibinfo{author}{Chen, X.}, \bibinfo{author}{Chen, L.}, \bibinfo{author}{Chen, L.}, \bibinfo{author}{Chen, P.}, \bibinfo{author}{Sheng, G.}, \bibinfo{author}{Yu, X.}, \bibinfo{author}{Zou, Y.}, \bibinfo{year}{2024}.
\newblock \bibinfo{title}{Modeling thermal infrared image degradation and real-world super-resolution under background thermal noise and streak interference}.
\newblock \bibinfo{journal}{IEEE Transactions on Circuits and Systems for Video Technology} \bibinfo{volume}{34}, \bibinfo{pages}{6194--6206}.
\newblock \DOIprefix\doi{10.1109/TCSVT.2023.3349182}.
\bibitem[{Chen et~al.(2023a)Chen, Wang, Zhou, Qiao and Dong}]{chen2023activating}
\bibinfo{author}{Chen, X.}, \bibinfo{author}{Wang, X.}, \bibinfo{author}{Zhou, J.}, \bibinfo{author}{Qiao, Y.}, \bibinfo{author}{Dong, C.}, \bibinfo{year}{2023}a.
\newblock \bibinfo{title}{Activating more pixels in image super-resolution transformer}, in: \bibinfo{booktitle}{Proceedings of the IEEE/CVF Conference on Computer Vision and Pattern Recognition}, pp. \bibinfo{pages}{22367--22377}.
\bibitem[{Chen et~al.(2023b)Chen, Tang, Zhang and Liu}]{chen2023model}
\bibinfo{author}{Chen, Y.}, \bibinfo{author}{Tang, Z.}, \bibinfo{author}{Zhang, K.}, \bibinfo{author}{Liu, Y.}, \bibinfo{year}{2023}b.
\newblock \bibinfo{title}{Model transferability with responsive decision subjects}, in: \bibinfo{booktitle}{International Conference on Machine Learning}, \bibinfo{organization}{PMLR}. pp. \bibinfo{pages}{4921--4952}.
\bibitem[{Dong et~al.(2015)Dong, Loy, He and Tang}]{dong2015image}
\bibinfo{author}{Dong, C.}, \bibinfo{author}{Loy, C.C.}, \bibinfo{author}{He, K.}, \bibinfo{author}{Tang, X.}, \bibinfo{year}{2015}.
\newblock \bibinfo{title}{Image super-resolution using deep convolutional networks}.
\newblock \bibinfo{journal}{IEEE transactions on pattern analysis and machine intelligence} \bibinfo{volume}{38}, \bibinfo{pages}{295--307}.
\bibitem[{Dong et~al.(2016)Dong, Loy and Tang}]{dong2016accelerating}
\bibinfo{author}{Dong, C.}, \bibinfo{author}{Loy, C.C.}, \bibinfo{author}{Tang, X.}, \bibinfo{year}{2016}.
\newblock \bibinfo{title}{Accelerating the super-resolution convolutional neural network}, in: \bibinfo{booktitle}{Computer Vision--ECCV 2016: 14th European Conference, Amsterdam, The Netherlands, October 11-14, 2016, Proceedings, Part II 14}, \bibinfo{organization}{Springer}. pp. \bibinfo{pages}{391--407}.
\bibitem[{Gupta and Mitra(2021)}]{gupta2021toward}
\bibinfo{author}{Gupta, H.}, \bibinfo{author}{Mitra, K.}, \bibinfo{year}{2021}.
\newblock \bibinfo{title}{Toward unaligned guided thermal super-resolution}.
\newblock \bibinfo{journal}{IEEE Transactions on Image Processing} \bibinfo{volume}{31}, \bibinfo{pages}{433--445}.
\bibitem[{He et~al.(2018)He, Tang, Yang, Cao, Yang and Cao}]{he2018cascaded}
\bibinfo{author}{He, Z.}, \bibinfo{author}{Tang, S.}, \bibinfo{author}{Yang, J.}, \bibinfo{author}{Cao, Y.}, \bibinfo{author}{Yang, M.Y.}, \bibinfo{author}{Cao, Y.}, \bibinfo{year}{2018}.
\newblock \bibinfo{title}{Cascaded deep networks with multiple receptive fields for infrared image super-resolution}.
\newblock \bibinfo{journal}{IEEE transactions on circuits and systems for video technology} \bibinfo{volume}{29}, \bibinfo{pages}{2310--2322}.
\bibitem[{Honda et~al.(2019)Honda, Sugimura and Hamamoto}]{honda2019multi}
\bibinfo{author}{Honda, T.}, \bibinfo{author}{Sugimura, D.}, \bibinfo{author}{Hamamoto, T.}, \bibinfo{year}{2019}.
\newblock \bibinfo{title}{Multi-frame rgb/nir imaging for low-light color image super-resolution}.
\newblock \bibinfo{journal}{IEEE Transactions on Computational Imaging} \bibinfo{volume}{6}, \bibinfo{pages}{248--262}.
\bibitem[{Huang et~al.(2021)Huang, Jiang, Lan, Zhang and Pi}]{huang2021infrared}
\bibinfo{author}{Huang, Y.}, \bibinfo{author}{Jiang, Z.}, \bibinfo{author}{Lan, R.}, \bibinfo{author}{Zhang, S.}, \bibinfo{author}{Pi, K.}, \bibinfo{year}{2021}.
\newblock \bibinfo{title}{Infrared image super-resolution via transfer learning and psrgan}.
\newblock \bibinfo{journal}{IEEE Signal Processing Letters} \bibinfo{volume}{28}, \bibinfo{pages}{982--986}.
\bibitem[{Huang et~al.(2022)Huang, Miyazaki, Liu and Omachi}]{huang2022infrared}
\bibinfo{author}{Huang, Y.}, \bibinfo{author}{Miyazaki, T.}, \bibinfo{author}{Liu, X.}, \bibinfo{author}{Omachi, S.}, \bibinfo{year}{2022}.
\newblock \bibinfo{title}{Infrared image super-resolution: Systematic review, and future trends}.
\newblock \bibinfo{journal}{arXiv preprint arXiv:2212.12322} .
\bibitem[{Jiang et~al.(2023a)Jiang, Li, Li, Li, Zhang and Lu}]{jiang2023enhanced}
\bibinfo{author}{Jiang, B.}, \bibinfo{author}{Li, J.}, \bibinfo{author}{Li, H.}, \bibinfo{author}{Li, R.}, \bibinfo{author}{Zhang, D.}, \bibinfo{author}{Lu, G.}, \bibinfo{year}{2023}a.
\newblock \bibinfo{title}{Enhanced frequency fusion network with dynamic hash attention for image denoising}.
\newblock \bibinfo{journal}{Information Fusion} \bibinfo{volume}{92}, \bibinfo{pages}{420--434}.
\bibitem[{Jiang et~al.(2023b)Jiang, Lu, Chen, Lu and Lu}]{jiang2023graph}
\bibinfo{author}{Jiang, B.}, \bibinfo{author}{Lu, Y.}, \bibinfo{author}{Chen, X.}, \bibinfo{author}{Lu, X.}, \bibinfo{author}{Lu, G.}, \bibinfo{year}{2023}b.
\newblock \bibinfo{title}{Graph attention in attention network for image denoising}.
\newblock \bibinfo{journal}{IEEE Transactions on Systems, Man, and Cybernetics: Systems} .
\bibitem[{Jiang et~al.(2022a)Jiang, Lu, Wang, Lu and Zhang}]{jiang2022deep}
\bibinfo{author}{Jiang, B.}, \bibinfo{author}{Lu, Y.}, \bibinfo{author}{Wang, J.}, \bibinfo{author}{Lu, G.}, \bibinfo{author}{Zhang, D.}, \bibinfo{year}{2022}a.
\newblock \bibinfo{title}{Deep image denoising with adaptive priors}.
\newblock \bibinfo{journal}{IEEE Transactions on Circuits and Systems for Video Technology} \bibinfo{volume}{32}, \bibinfo{pages}{5124--5136}.
\bibitem[{Jiang et~al.(2023c)Jiang, Lu, Zhang and Lu}]{jiang2023few}
\bibinfo{author}{Jiang, B.}, \bibinfo{author}{Lu, Y.}, \bibinfo{author}{Zhang, B.}, \bibinfo{author}{Lu, G.}, \bibinfo{year}{2023}c.
\newblock \bibinfo{title}{Few-shot learning for image denoising}.
\newblock \bibinfo{journal}{IEEE Transactions on Circuits and Systems for Video Technology} \bibinfo{volume}{33}, \bibinfo{pages}{4741--4753}.
\bibitem[{Jiang et~al.(2024)Jiang, Lu, Zhang and Lu}]{10314035}
\bibinfo{author}{Jiang, B.}, \bibinfo{author}{Lu, Y.}, \bibinfo{author}{Zhang, B.}, \bibinfo{author}{Lu, G.}, \bibinfo{year}{2024}.
\newblock \bibinfo{title}{Agp-net: Adaptive graph prior network for image denoising}.
\newblock \bibinfo{journal}{IEEE Transactions on Industrial Informatics} \bibinfo{volume}{20}, \bibinfo{pages}{4753--4764}.
\newblock \DOIprefix\doi{10.1109/TII.2023.3316184}.
\bibitem[{Jiang et~al.(2022b)Jiang, Wang, Lu, Lu and Zhang}]{jiang2022multilevel}
\bibinfo{author}{Jiang, B.}, \bibinfo{author}{Wang, J.}, \bibinfo{author}{Lu, Y.}, \bibinfo{author}{Lu, G.}, \bibinfo{author}{Zhang, D.}, \bibinfo{year}{2022}b.
\newblock \bibinfo{title}{Multilevel noise contrastive network for few-shot image denoising}.
\newblock \bibinfo{journal}{IEEE Transactions on Instrumentation and Measurement} \bibinfo{volume}{71}, \bibinfo{pages}{1--13}.
\bibitem[{Jing et~al.(2024)Jing, Shao and Zhong}]{jing2024boosting}
\bibinfo{author}{Jing, D.}, \bibinfo{author}{Shao, H.}, \bibinfo{author}{Zhong, D.}, \bibinfo{year}{2024}.
\newblock \bibinfo{title}{Boosting edge detection via fusing spatial and frequency domains}.
\newblock \bibinfo{journal}{Pattern Recognition} , \bibinfo{pages}{110699}.
\bibitem[{Kanopoulos et~al.(1988)Kanopoulos, Vasanthavada and Baker}]{kanopoulos1988design}
\bibinfo{author}{Kanopoulos, N.}, \bibinfo{author}{Vasanthavada, N.}, \bibinfo{author}{Baker, R.L.}, \bibinfo{year}{1988}.
\newblock \bibinfo{title}{Design of an image edge detection filter using the sobel operator}.
\newblock \bibinfo{journal}{IEEE Journal of solid-state circuits} \bibinfo{volume}{23}, \bibinfo{pages}{358--367}.
\bibitem[{Ledig et~al.(2017)Ledig, Theis, Husz{\'a}r, Caballero, Cunningham, Acosta, Aitken, Tejani, Totz, Wang et~al.}]{ledig2017photo}
\bibinfo{author}{Ledig, C.}, \bibinfo{author}{Theis, L.}, \bibinfo{author}{Husz{\'a}r, F.}, \bibinfo{author}{Caballero, J.}, \bibinfo{author}{Cunningham, A.}, \bibinfo{author}{Acosta, A.}, \bibinfo{author}{Aitken, A.}, \bibinfo{author}{Tejani, A.}, \bibinfo{author}{Totz, J.}, \bibinfo{author}{Wang, Z.}, et~al., \bibinfo{year}{2017}.
\newblock \bibinfo{title}{Photo-realistic single image super-resolution using a generative adversarial network}, in: \bibinfo{booktitle}{Proceedings of the IEEE conference on computer vision and pattern recognition}, pp. \bibinfo{pages}{4681--4690}.
\bibitem[{Liang et~al.(2021)Liang, Cao, Sun, Zhang, Van~Gool and Timofte}]{liang2021swinir}
\bibinfo{author}{Liang, J.}, \bibinfo{author}{Cao, J.}, \bibinfo{author}{Sun, G.}, \bibinfo{author}{Zhang, K.}, \bibinfo{author}{Van~Gool, L.}, \bibinfo{author}{Timofte, R.}, \bibinfo{year}{2021}.
\newblock \bibinfo{title}{Swinir: Image restoration using swin transformer}, in: \bibinfo{booktitle}{Proceedings of the IEEE/CVF international conference on computer vision}, pp. \bibinfo{pages}{1833--1844}.
\bibitem[{Lim et~al.(2017)Lim, Son, Kim, Nah and Mu~Lee}]{lim2017enhanced}
\bibinfo{author}{Lim, B.}, \bibinfo{author}{Son, S.}, \bibinfo{author}{Kim, H.}, \bibinfo{author}{Nah, S.}, \bibinfo{author}{Mu~Lee, K.}, \bibinfo{year}{2017}.
\newblock \bibinfo{title}{Enhanced deep residual networks for single image super-resolution}, in: \bibinfo{booktitle}{Proceedings of the IEEE conference on computer vision and pattern recognition workshops}, pp. \bibinfo{pages}{136--144}.
\bibitem[{Liu et~al.(2022)Liu, Fan, Huang, Wu, Liu, Zhong and Luo}]{liu2022target}
\bibinfo{author}{Liu, J.}, \bibinfo{author}{Fan, X.}, \bibinfo{author}{Huang, Z.}, \bibinfo{author}{Wu, G.}, \bibinfo{author}{Liu, R.}, \bibinfo{author}{Zhong, W.}, \bibinfo{author}{Luo, Z.}, \bibinfo{year}{2022}.
\newblock \bibinfo{title}{Target-aware dual adversarial learning and a multi-scenario multi-modality benchmark to fuse infrared and visible for object detection}, in: \bibinfo{booktitle}{Proceedings of the IEEE/CVF Conference on Computer Vision and Pattern Recognition}, pp. \bibinfo{pages}{5802--5811}.
\bibitem[{Liu et~al.(2024)Liu, Li, Chen, Rao, Zuo, Wang, Yan, Wang, Chen and Lv}]{liu2024dsfusion}
\bibinfo{author}{Liu, K.}, \bibinfo{author}{Li, M.}, \bibinfo{author}{Chen, C.}, \bibinfo{author}{Rao, C.}, \bibinfo{author}{Zuo, E.}, \bibinfo{author}{Wang, Y.}, \bibinfo{author}{Yan, Z.}, \bibinfo{author}{Wang, B.}, \bibinfo{author}{Chen, C.}, \bibinfo{author}{Lv, X.}, \bibinfo{year}{2024}.
\newblock \bibinfo{title}{Dsfusion: Infrared and visible image fusion method combining detail and scene information}.
\newblock \bibinfo{journal}{Pattern Recognition} , \bibinfo{pages}{110633}.
\bibitem[{Liu et~al.(2023)Liu, Yin, Yang, Wang and An}]{liu2023combining}
\bibinfo{author}{Liu, T.}, \bibinfo{author}{Yin, Q.}, \bibinfo{author}{Yang, J.}, \bibinfo{author}{Wang, Y.}, \bibinfo{author}{An, W.}, \bibinfo{year}{2023}.
\newblock \bibinfo{title}{Combining deep denoiser and low-rank priors for infrared small target detection}.
\newblock \bibinfo{journal}{Pattern Recognition} \bibinfo{volume}{135}, \bibinfo{pages}{109184}.
\bibitem[{Lukose et~al.(2021)Lukose, Chidangil and George}]{Lukose2021OpticalTF}
\bibinfo{author}{Lukose, J.}, \bibinfo{author}{Chidangil, S.}, \bibinfo{author}{George, S.D.}, \bibinfo{year}{2021}.
\newblock \bibinfo{title}{Optical technologies for the detection of viruses like covid-19: Progress and prospects}.
\newblock \bibinfo{journal}{Biosensors \& Bioelectronics} \bibinfo{volume}{178}, \bibinfo{pages}{113004 -- 113004}.
\bibitem[{Nandhakumar et~al.(1994)Nandhakumar, Karthik and Aggarwal}]{nandhakumar1994unified}
\bibinfo{author}{Nandhakumar, N.}, \bibinfo{author}{Karthik, S.}, \bibinfo{author}{Aggarwal, J.K.}, \bibinfo{year}{1994}.
\newblock \bibinfo{title}{Unified modeling of non-homogeneous 3d objects for thermal and visual image synthesis}.
\newblock \bibinfo{journal}{Pattern Recognition} \bibinfo{volume}{27}, \bibinfo{pages}{1303--1316}.
\bibitem[{Peng et~al.(2019)Peng, Huang, Sun and Saenko}]{peng2019domain}
\bibinfo{author}{Peng, X.}, \bibinfo{author}{Huang, Z.}, \bibinfo{author}{Sun, X.}, \bibinfo{author}{Saenko, K.}, \bibinfo{year}{2019}.
\newblock \bibinfo{title}{Domain agnostic learning with disentangled representations}, in: \bibinfo{booktitle}{International Conference on Machine Learning}, \bibinfo{organization}{PMLR}. pp. \bibinfo{pages}{5102--5112}.
\bibitem[{Sengupta et~al.(2019)Sengupta, Ye, Wang, Liu and Roy}]{sengupta2019going}
\bibinfo{author}{Sengupta, A.}, \bibinfo{author}{Ye, Y.}, \bibinfo{author}{Wang, R.}, \bibinfo{author}{Liu, C.}, \bibinfo{author}{Roy, K.}, \bibinfo{year}{2019}.
\newblock \bibinfo{title}{Going deeper in spiking neural networks: Vgg and residual architectures}.
\newblock \bibinfo{journal}{Frontiers in neuroscience} \bibinfo{volume}{13}, \bibinfo{pages}{95}.
\bibitem[{Shi et~al.(2023)Shi, Wu, Han, Shao, Li and Wu}]{shi2023source}
\bibinfo{author}{Shi, Y.}, \bibinfo{author}{Wu, K.}, \bibinfo{author}{Han, Y.}, \bibinfo{author}{Shao, Y.}, \bibinfo{author}{Li, B.}, \bibinfo{author}{Wu, F.}, \bibinfo{year}{2023}.
\newblock \bibinfo{title}{Source-free and black-box domain adaptation via distributionally adversarial training}.
\newblock \bibinfo{journal}{Pattern Recognition} \bibinfo{volume}{143}, \bibinfo{pages}{109750}.
\bibitem[{Song et~al.(2019)Song, Deng, Mota, Deligiannis, Dragotti and Rodrigues}]{song2019multimodal}
\bibinfo{author}{Song, P.}, \bibinfo{author}{Deng, X.}, \bibinfo{author}{Mota, J.F.}, \bibinfo{author}{Deligiannis, N.}, \bibinfo{author}{Dragotti, P.L.}, \bibinfo{author}{Rodrigues, M.R.}, \bibinfo{year}{2019}.
\newblock \bibinfo{title}{Multimodal image super-resolution via joint sparse representations induced by coupled dictionaries}.
\newblock \bibinfo{journal}{IEEE Transactions on Computational Imaging} \bibinfo{volume}{6}, \bibinfo{pages}{57--72}.
\bibitem[{Su{\'a}rez et~al.(2024)Su{\'a}rez, Carpio and Sappa}]{suarez2024enhancement}
\bibinfo{author}{Su{\'a}rez, P.L.}, \bibinfo{author}{Carpio, D.}, \bibinfo{author}{Sappa, A.D.}, \bibinfo{year}{2024}.
\newblock \bibinfo{title}{Enhancement of guided thermal image super-resolution approaches}.
\newblock \bibinfo{journal}{Neurocomputing} \bibinfo{volume}{573}, \bibinfo{pages}{127197}.
\bibitem[{Sun et~al.(2016)Sun, Feng and Saenko}]{sun2016return}
\bibinfo{author}{Sun, B.}, \bibinfo{author}{Feng, J.}, \bibinfo{author}{Saenko, K.}, \bibinfo{year}{2016}.
\newblock \bibinfo{title}{Return of frustratingly easy domain adaptation}, in: \bibinfo{booktitle}{Proceedings of the AAAI conference on artificial intelligence}.
\bibitem[{Sun et~al.(2024)Sun, Luo, Ren, Du, Chang and Wan}]{sun2024unsupervised}
\bibinfo{author}{Sun, H.}, \bibinfo{author}{Luo, Z.}, \bibinfo{author}{Ren, D.}, \bibinfo{author}{Du, B.}, \bibinfo{author}{Chang, L.}, \bibinfo{author}{Wan, J.}, \bibinfo{year}{2024}.
\newblock \bibinfo{title}{Unsupervised multi-branch network with high-frequency enhancement for image dehazing}.
\newblock \bibinfo{journal}{Pattern Recognition} \bibinfo{volume}{156}, \bibinfo{pages}{110763}.
\bibitem[{Sun et~al.(2023)Sun, Dong, Tang and Pan}]{sun2023spatially}
\bibinfo{author}{Sun, L.}, \bibinfo{author}{Dong, J.}, \bibinfo{author}{Tang, J.}, \bibinfo{author}{Pan, J.}, \bibinfo{year}{2023}.
\newblock \bibinfo{title}{Spatially-adaptive feature modulation for efficient image super-resolution}, in: \bibinfo{booktitle}{Proceedings of the IEEE/CVF International Conference on Computer Vision}, pp. \bibinfo{pages}{13190--13199}.
\bibitem[{Sun et~al.(2022)Sun, Pan and Tang}]{sun2022shufflemixer}
\bibinfo{author}{Sun, L.}, \bibinfo{author}{Pan, J.}, \bibinfo{author}{Tang, J.}, \bibinfo{year}{2022}.
\newblock \bibinfo{title}{Shufflemixer: An efficient convnet for image super-resolution}.
\newblock \bibinfo{journal}{Advances in Neural Information Processing Systems} \bibinfo{volume}{35}, \bibinfo{pages}{17314--17326}.
\bibitem[{Wang et~al.(2023)Wang, Zhu, Zhang and Wang}]{wang2023m}
\bibinfo{author}{Wang, P.}, \bibinfo{author}{Zhu, H.}, \bibinfo{author}{Zhang, H.}, \bibinfo{author}{Wang, N.}, \bibinfo{year}{2023}.
\newblock \bibinfo{title}{M-cbn: Manifold constrained joint image dehazing and super-resolution based on chord boosting network}.
\newblock \bibinfo{journal}{Pattern Recognition} \bibinfo{volume}{135}, \bibinfo{pages}{109166}.
\bibitem[{Wang et~al.(2021)Wang, Xie, Dong and Shan}]{wang2021real}
\bibinfo{author}{Wang, X.}, \bibinfo{author}{Xie, L.}, \bibinfo{author}{Dong, C.}, \bibinfo{author}{Shan, Y.}, \bibinfo{year}{2021}.
\newblock \bibinfo{title}{Real-esrgan: Training real-world blind super-resolution with pure synthetic data}, in: \bibinfo{booktitle}{Proceedings of the IEEE/CVF international conference on computer vision}, pp. \bibinfo{pages}{1905--1914}.
\bibitem[{Wang et~al.(2018)Wang, Yu, Wu, Gu, Liu, Dong, Qiao and Change~Loy}]{wang2018esrgan}
\bibinfo{author}{Wang, X.}, \bibinfo{author}{Yu, K.}, \bibinfo{author}{Wu, S.}, \bibinfo{author}{Gu, J.}, \bibinfo{author}{Liu, Y.}, \bibinfo{author}{Dong, C.}, \bibinfo{author}{Qiao, Y.}, \bibinfo{author}{Change~Loy, C.}, \bibinfo{year}{2018}.
\newblock \bibinfo{title}{Esrgan: Enhanced super-resolution generative adversarial networks}, in: \bibinfo{booktitle}{Proceedings of the European conference on computer vision (ECCV) workshops}, pp. \bibinfo{pages}{0--0}.
\bibitem[{Wang et~al.(2020)Wang, Chen and Hoi}]{wang2020deep}
\bibinfo{author}{Wang, Z.}, \bibinfo{author}{Chen, J.}, \bibinfo{author}{Hoi, S.C.}, \bibinfo{year}{2020}.
\newblock \bibinfo{title}{Deep learning for image super-resolution: A survey}.
\newblock \bibinfo{journal}{IEEE transactions on pattern analysis and machine intelligence} \bibinfo{volume}{43}, \bibinfo{pages}{3365--3387}.
\bibitem[{Wei et~al.(2021)Wei, Lan, Zeng, Zhang and Chen}]{wei2021toalign}
\bibinfo{author}{Wei, G.}, \bibinfo{author}{Lan, C.}, \bibinfo{author}{Zeng, W.}, \bibinfo{author}{Zhang, Z.}, \bibinfo{author}{Chen, Z.}, \bibinfo{year}{2021}.
\newblock \bibinfo{title}{Toalign: task-oriented alignment for unsupervised domain adaptation}.
\newblock \bibinfo{journal}{Advances in Neural Information Processing Systems} \bibinfo{volume}{34}, \bibinfo{pages}{13834--13846}.
\bibitem[{Wei and Yuan(2023)}]{wei2023adversarial}
\bibinfo{author}{Wei, X.}, \bibinfo{author}{Yuan, M.}, \bibinfo{year}{2023}.
\newblock \bibinfo{title}{Adversarial pan-sharpening attacks for object detection in remote sensing}.
\newblock \bibinfo{journal}{Pattern Recognition} \bibinfo{volume}{139}, \bibinfo{pages}{109466}.
\bibitem[{Yu et~al.(1998)Yu, Peng, Tu and Wang}]{yu1998infrared}
\bibinfo{author}{Yu, W.}, \bibinfo{author}{Peng, Q.}, \bibinfo{author}{Tu, H.}, \bibinfo{author}{Wang, Z.}, \bibinfo{year}{1998}.
\newblock \bibinfo{title}{An infrared image synthesis model based on infrared physics and heat transfer}.
\newblock \bibinfo{journal}{International journal of infrared and millimeter waves} \bibinfo{volume}{19}, \bibinfo{pages}{1661--1669}.
\bibitem[{Zhang et~al.(2018)Zhang, Li, Li, Wang, Zhong and Fu}]{zhang2018image}
\bibinfo{author}{Zhang, Y.}, \bibinfo{author}{Li, K.}, \bibinfo{author}{Li, K.}, \bibinfo{author}{Wang, L.}, \bibinfo{author}{Zhong, B.}, \bibinfo{author}{Fu, Y.}, \bibinfo{year}{2018}.
\newblock \bibinfo{title}{Image super-resolution using very deep residual channel attention networks}, in: \bibinfo{booktitle}{Proceedings of the European conference on computer vision (ECCV)}, pp. \bibinfo{pages}{286--301}.
\bibitem[{Zhao et~al.(2024)Zhao, Gao, Deng and Li}]{zhao2024ssir}
\bibinfo{author}{Zhao, L.}, \bibinfo{author}{Gao, J.}, \bibinfo{author}{Deng, D.}, \bibinfo{author}{Li, X.}, \bibinfo{year}{2024}.
\newblock \bibinfo{title}{Ssir: Spatial shuffle multi-head self-attention for single image super-resolution}.
\newblock \bibinfo{journal}{Pattern Recognition} \bibinfo{volume}{148}, \bibinfo{pages}{110195}.

\end{thebibliography}

\end{document}